\documentclass[aps,groupedaddress,showpacs]{revtex4} 
\usepackage{graphicx}

\begin{document}

% You should use BibTeX and revtex.bst for references 
\bibliographystyle{revtex} 
% marks overfull lines with blackboxes 
\draft

% Use the \preprint command to place your local institutional report 
% number on the title page in preprint mode.  
% Multiple \preprint commands are allowed.  
%\preprint{}

%Title of paper 
\title{Onset of the nonlinear dielectric response of glasses in the two-level system model.} 
% Optional argument for running titles on pages 
%\title[]{}

% repeat the \author ..  \affiliation etc.  as needed 
% \email, \thanks, \homepage, \altaffiliation all apply to the current 
% author. Explanatory text should go in the []'s, actual e-mail 
% address or url should go in the {}'s for \email and \homepage.  
% Please use the appropriate macro for the type of information

% \affiliation command applies to all authors since the last 
% \affiliation command.  The \affiliation command should follow the 
% other information

\author{J.  Le Cochec} 
\author{F.  Ladieu} 
\email[]{ladieu@drecam.cea.fr} 
\affiliation{DSM/DRECAM/LPS, C.E.Saclay, 91191 Gif sur Yvette Cedex, France}

\date{\today}

\begin{abstract} 
We have calculated the real part $\chi'$ of the nonlinear dielectric susceptibility of amorphous insulators in the kHz 
range, 
by using the two-level system 
model and a nonperturbative numerical quantum approach. At low temperature $T$, it is first shown that the standard 
two-level 
model should lead to a \textit{decrease} of $\chi'$ when the measuring field $E$ is raised, 
since raising $E$ increases the population of the upper level and induces Rabi oscillations cancelling the ones induced 
from 
the ground level. This predicted $E$-induced decrease of $\chi'$ is at \textit{odds} with experiments. However, a
\textit{good agreement} with low-frequency experimental nonlinear data is achieved if, in our fully quantum simulations, 
interactions between 
defects are taken into
account by a new relaxation rate whose efficiency increases as $\sqrt{E}$, as was proposed recently by Burin \textit{et 
al.} 
(Phys. Rev. Lett. {\bf 86}, 5616 (2001)).  In this approach, the behavior of $\chi'$ at low $T$ is mainly explained by the 
efficiency of this new relaxation channel. This new relaxation rate could be further tested since it is shown that it 
should 
lead: \textit{ i)} to a completely new nonlinear behavior for samples whose thickness is $\simeq 10$ nm; \textit{ ii)} to a 
decrease of nonequilibrium effects when $E$ is increased.  
\end{abstract}

% insert suggested PACS numbers in braces on next line 

%\maketitle must follow title, authors, abstract and \pacs 
\maketitle 
% body of paper here - Use proper section commands 
% References should be done using the \cite, \ref, and \label commands 

Amorphous materials exhibit universal anomalous properties at low temperature. In 1971, Zeller and Pohl \cite{Zeller} 
discovered below 1 K a quasilinear behavior of the specific heat in a number of glasses contrasting with the Debye law of 
crystalline materials. Anderson, Halperin, Varma \cite{Anderson} and Phillips \cite{Phillips72} proposed an explaination 
based 
upon the existence of localized two-level systems (TLS). Their origin may be due to the tunneling of atoms or groups of 
atoms 
between two equilibrium positions separated by a narrow energy barrier featuring asymmetric two-well potentials.  They are 
assumed randomly distributed in energy splittings and tunneling barriers as a consequence of the structural disorder of 
these 
materials. This model has proven to be successful to understand most salient experimental features.

The standard TLS model assumes defects do not interact with one another. However, defects are strongly coupled to their 
environment and can emit or absorb phonons. It leads to an indirect interaction between nearest neighbors via the phonon 
field 
\cite{Joffrin}. Certain recent failures to explain nonequilibrium data (at a few $\mathrm{kHz}$) \cite{Caruzzo} underscore 
the 
likely involvement of these interactions below $100$ $\mathrm{mK}$. However, these nonequilibrium effects are small 
corrections of the $\mathrm{kHz}$ stationnary response, and, up to recently, examples of stationnary susceptibilities 
strongly 
affected by interactions were very rare : in the $\mathrm{kHz}$ regime, it was argued that the ultra-low-T ($T\simeq 1$ 
$\mathrm{mK}$) plateau of the dielectric constant in the linear regime, strongly different from the expected logarithmic 
increase, resulted from interactions \cite{Enss}. Very recently, such a conclusion was drawn from internal friction 
experiments \cite{Thompson}.

In this work, we show that including interactions in the TLS model with a recently proposed mechanism \cite{Burin} 
\textit{strongly} affects the
 nonlinear \textit{stationnary} dielectric susceptibility $\chi'$ of amorphous insulators at a few $\mathrm{kHz}$. A very 
complete set of such data was published a few years ago by Rogge \textit{et al.} \cite{Rogge}, twenty years after the 
pioneering work of Frossati \textit{et al.} \cite{Frossati}. In the linear regime, $\chi'$ decreases when $T$ decreases, 
reaches its 
minimum at $T_{rev}$ and then increases below $T_{rev}$ (before reaching the  
above-mentioned ultra-low-$T$ plateau $\chi'_{plat}$). According to the standard TLS 
model, the $\chi'$ decrease above $T_{rev}$ is due to the progressive freezing of 
the diagonal (or relaxational) part of the response, while the $\chi'$ 
increase below $T_{rev}$ comes 
from the induced off-diagonal (or resonant) part of the susceptibility : 
this effect enlarges as $T$ decreases as do all quantum effects. However, due to
 the quantum nature of $\chi'$ below $T_{rev}$, one expects $\chi'$ to be
 strongly \textit{depressed} by a strong measuring 
electric field $E$ at a given $T$.
 This can be guessed from the \textit{quantum saturation} 
phenomenon which is \textit{very general} in two level systems \cite{AtomPhoton}. Indeed, increasing $E$ decreases the 
population 
difference between the two energy levels : as the Rabi oscillations produced by 
$E$ on the upper level are in phase opposition with respect to the ones produced on the ground level, the quantum response, 
once averaged on many independent TLS's, tends to zero when $E$ is increased. Strikingly, Rogge \textit{et al.} experiments 
show the opposite trend : $\chi'(T<T_{rev})$ 
\textit{ increases} when $E$ is increased. 

As it is carefully explained in Ref. \cite{Rogge}, this behavior does \textit{not} result from heating of the 
sample by $E$. To give a supplementary argument with respect to Ref. \cite{Rogge}, let us note that if $E_{lin}$ is the 
upper 
field below which the dielectric susceptibility is measured as being field 
independent, one expects that the heating of the sample, for a given $E \gtrsim E_{lin}$, is more important when $T$ 
decreases. A heating effect is thus expected to strech the $\chi'(T)$ curve of an amount \textit{increasing} as $T$ 
decreases, 
i.e. one expects

 $$\left\vert \frac{\partial \chi'}{\partial T} \right\vert_{E \gtrsim E_{lin}} < \left\vert \frac{\partial \chi'}{\partial 
T} 
\right\vert_{E 
\leq E_{lin}}, \eqno(1)$$ 

to hold at low $T$, i.e. mainly at $T\leq T_{rev.}$. As can be seen, e.g. on the Fig. 3 of Ref. \cite{Rogge}, the trend of 
the 
data 
is \textit{exactly the opposite} of Eq. (1). Finally, since heating effects can be ruled out, the fact that  below 
$T_{rev}$, 
$\chi'(E \gtrsim E_{lin})$ does not behave as expected from the quantum saturation phenomenon seems extremely intriguing 
 in the framework of the standard TLS model.

However, this was not
 pointed out since the non linear effects in the TLS model were, up to now, only calculated by using \textit{the adiabatic 
approximation} \cite{Stockburger}. Such an approximation  
states that TLS's are at every moment at thermal equilibrium, i.e., it disregards 
any coherence effects.  It predicts an increase of $\chi'$ with $E$, i.e. it qualitatively accounts for the experimental 
behavior. However, in the specific case of the real part of the susceptibility, the consistency of the adiabatic 
approximation is questionnable \cite{noteadia}. Indeed, as it is very clearly stated in Ref. \cite{Stockburger}, this approximation does not hold 
for TLS's whose Tunneling energy $\Delta_0$ is too small, and yet it finds that the nonlinear part of $\chi'$ is dominated by the 
smallest $\Delta_0$ values (see after Eq. (3.30) in Ref. \cite{Stockburger}). More precisely \cite{Stockburger}, with $p_0 \simeq 
1$ D the TLS dipole, even for the lowest electric fields $E \simeq 1$ kV/m of frequency $\omega \simeq 1$ kHz, the adiabatic 
approximation fails when $\Delta_0 \lesssim \sqrt{ \hbar \omega p_0 E} \simeq 3$ $\mu$K, while it is well known, from 
instationnarity experiments \cite{Caruzzo}, that smaller Tunneling energies exist in glasses. Besides, the second puzzling point is 
that, according to the authors 
themselves \cite{Stockburger}, the reason
 of the increase of $\chi'$ with $E$ in the adiabatic approximation is
 physically obscure, which leaves unsolved the question of the expected "quantum saturation effect" above mentionned. Finally, 
several predictions of Ref. \cite{Stockburger} are somehow contradicted by experiments \cite{Rogge} : instead of the predicted 
$T_{rev} \propto E^\gamma$ with $\gamma > 1$,  the measured data yield $\gamma \lesssim 1/2$; below $T_{rev}$, at a given $E$, the 
predicted peaked behavior of $\partial \chi'/ \partial T$ is not observed; at very low $T$, the observed $E$ dependence of 
$\chi'_{plat}$ contradicts the predictions. 
 
  This work goes beyond the adiabatic approximation, even though, due to the few simplifying assumptions that we have made (see Eq. 
(2)), we do not intend to yield a fully "from first principle 
calculation". The key point is that phase coherence is not discarded here since non linear 
effects are treated by a fully quantum non perturbative method. In the first part, we show that the standard TLS model cannot 
explain the low-frequency experimental 
data 
below $100$ $\mathrm{mK}$ since it yields, at low $T$,  the above-mentioned quantum saturation phenomenon. In a second 
part, 
interactions between defects are added by using an  interaction 
mechanism proposed very recently by Burin \textit{et al.} \cite{Burin}, and a successful agreement is obtained with 
experiments. Finally, we briefly discuss experimental predictions implied by Burin \textit{et al.}'s interaction mechanism. 

\section{Standard two-level system model }

\subsection{Bloch equations of TLS}

\subsubsection{Dynamics of a unique isolated TLS}

Consider a TLS that is sitting in a double-well potential and assume this defect has a dipole moment $\mathbf{p_0}$. Its 
energy splitting $\epsilon$ is related \cite{Phillips} to the asymmetry energy $\Delta$ and to the tunneling matrix element 
$\Delta_{0}$, 
describing transitions between the wells, by $\epsilon=\sqrt{\Delta^2+\Delta_0^2}$. Due to finite 
$\Delta_{0}$, the eigen states extend over both sides of the TLS, and the position operator $\mathbf{r}$
 is no longer diagonal in this eigen basis. As a result, when an external electric field $\mathbf{E}$ is applied to 
$\mathbf{p_0}$, the coupling Hamiltonian $q \mathbf{E}.\mathbf{r}$ is not diagonal in the eigen basis \cite{Caruzzo} (upon 
which all the operators of this work are expressed), yielding a total Hamiltonian :

$$H = \frac{1}{2} \left( \begin{array}{cc}\epsilon&0\\0&-\epsilon\end{array} \right) + \left( 
\begin{array}{cc}\frac{\Delta}{\epsilon}&\frac{\Delta_0}{\epsilon}  \\ 
\frac{\Delta_0}{\epsilon}&-\frac{\Delta}{\epsilon}\end{array} \right) \mathbf{p_0}.\mathbf{E}\cos\omega t , $$
  
or $H= -\mathbf{s}.\mathbf{\Omega}$, with $\mathbf{s}=\frac{\hbar}{2} \mathbf{\Sigma}$ where $\mathbf{\Sigma}$ are the 
three 
Pauli matrices and $\mathbf{\Omega }$ is an external effective field ($\mathbf{\Omega}$ components are given below, note 
$\Omega_y=0$), which shows an effective spin operator $\mathbf{s}$ is associated to the TLS. The systematic use of "spin" 
language 
comes from the fact that the three Pauli matrices, combined with the identity matrix, form a general basis for TLS's. 
Whatever 
its physical nature, any operator can be expressed as a linear combination of these four matrices, e.g., the density 
operator 
$\rho$ can be written : $ \rho= (1/2)I + (1/\hbar){\mathbf{S}}.\mathbf{\Sigma}$, where $\mathbf{S}$ is the quantum mean 
value of the 
spin operator $\mathbf{s}$ . This shows that  ${S}_x$ and 
${S}_y$ describe the coherence effects contained in the off-diagonal terms of $\rho$, while ${S}_z$ is proportional to 
the population difference 
between the levels (the occupation probabilities are given by the diagonal terms of $\rho$).

The movement of $\mathbf{p_0}$ and thereafter the dielectric response of the material stem from the dynamics of 
$\mathbf{S}$. 
For a perfectly isolated TLS (note that this implies that $T=0$) the evolution of $\mathbf{S}$ in the external field 
$\mathbf{\Omega}$ is only a precession around the external field $\mathbf{\Omega}$, as can be seen from the Schr\"odinger 
equation which leads \cite{Caruzzo} to $\partial \mathbf{S}/\partial t = \mathbf{S} \times \mathbf{\Omega}$.

\subsubsection{Dynamics of an ensemble of non-isolated TLS's}

At finite $T$, the dynamics of the TLS must include the relaxation toward its
 equilibrium value since each TLS interacts with its environment (phonons or neighboring defects). Since these interactions 
occur randomly for a given TLS, the dynamical equation must deal with ensemble averaged properties $\bar{\mathbf{S}}$, 
i.e. with quantities averaged over many similar TLS's. This evolution is given by the Bloch equations, namely 
 
$$\frac{\partial \bar{\mathbf{S}} }{\partial t} = \bar{\mathbf{S}} \times \mathbf{\Omega} + \frac{ 
{\bar{\mathbf{S}} } - <\bar{\mathbf{S}}>_{relax} }{\tau_{relax}}, \eqno(2)$$

where the last term states that the relaxation of $\bar{\mathbf{S}}$ toward the environement equilibrium values 
$<\bar{\mathbf{S}}>_{relax}$ must be added to the quantum dynamics (see Appendix {\bf A}).  In Eq. (2) it is assumed that 
the relaxation of a given $\bar{\mathbf{S}}$ component, say $\bar{S}_{x}$, occurs with a well defined time constant, say 
$\tau_x$. In the important case of short time scales, one needs to go beyond this approximation since echo signals do 
not generally decay as a simple exponential (\cite{Laikhtman}, \cite{Galperin}). This subtle effect is irrelevant here since, 
as already stated e.g. in Ref. \cite{Caruzzo}, we are 
only interested in the \textit{long time range} solution of Eq. (2), namely $\chi'(1 \mathrm{kHz})$, i.e. we focus on the 
particular case $\omega \tau_2 \ll 1$ (see below). Similarly the relaxation term of Eq. (2) might become more complicated in 
the case of very strong 
fields \cite{Geva}, leading, e.g., to a $\bar{S}_y/\tau_{x,y}$ 
term in the relaxation of $\bar{S}_x$ (see Appendix {\bf B}). However this should not be the case here 
 since we only focus on the \textit{onset} of the non linear regime ($p_0E$ will not much exceed $k_{B}T$). As a 
result, the relaxation 
terms can be derived quite simply, as we show now.

\vskip0.5cm

\textit{i)} Phonon induced relaxation.

Let us first focus on phonon field
 relaxation. The occupation probabilities are altered by the emission or the 
absorption of phonons, yielding \cite{Phillips} a relaxation of $\bar{S}_z$, with the relaxation
 time $\tau_1= \kappa_{1}/(\epsilon\Delta_0^2)\tanh{\frac{\epsilon}{2k_BT}}$ 
, where $\kappa_{1}$ is a sample-dependent constant. Since phonon  processes occur 
randomly and independently for various TLS's, they break the phase coherence of the ensemble of (noninteracting) TLS's, 
yielding a relaxation time $2\tau_{1}$ for $\bar{S}_{x}$ and 
$\bar{S}_{y}$. What are the thermodynamic values $<\bar{S}_{x,y,z}>$ to which $\bar{S}_{x,y,z}$ relax ? By second order 
expansion of 
dynamical correlation functions, it was shown 
\cite{Abragam} that this relaxation occurs towards the so-called "instantaneous equilibrium values", namely, 
$<\bar{S}_{x,y,z}(t)> = 
Tr(<\rho\left(t\right)> \bar{S}_{x,y,z})$ where $<\rho(t)> = \exp(-H(t)/(k_{B}T))/Tr(\exp(-H(t)/(k_{B}T))$ is the 
"instantaneous" 
thermodynamical density operator and $k_{B}$ is Boltzmann's constant. For this result to be valid, several conditions must 
be 
fullfilled, among which the most stringent one is,  by far : $\left|\mathbf{p_0}.\mathbf{E}\right| \tau_{c} \le \hbar$ 
where 
$\tau_{c}$ is the correlation time of the random electrical field acting on a given TLS due to its small interactions with 
its 
neighbors (see next paragraph \textit{ii)} ).
Finally, these phonon 
processes yield in the Bloch equations 
a term $(\bar{S}_{z}(t)-<\bar{S}_{z}(t)>)/\tau_{1}$ for the population relaxation, and  
$(\bar{S}_{x,y}(t)-<\bar{S}_{x,y}(t)>)/(2\tau_{1})$ for the relaxation of the coherence terms.
 
Does $\tau_1$ depend on time ? On one hand, under the above stated assumption $\left|\mathbf{p_0}.\mathbf{E}\right| 
\tau_{c} \le \hbar$, it was argued \cite{Abragam} that $\tau_{1}$ does not depend on time (see also Ref. \cite{t1t}). On the 
other hand, one may argue \cite{Stockburger} that, since the applied electric field modulates the asymetry energy $\Delta$, 
one should use $\tau_1 (t) = \kappa_{1}/(\epsilon_{eff}\Delta_0^2)\tanh{\frac{\epsilon_{eff}}{2k_BT}}$, where $\Delta_{eff} 
= \Delta + p_0E \cos{\omega t}$ and $\epsilon_{eff} = \sqrt{\Delta_{eff}^2 + \Delta_0^2}$ arise from the diagonalisation of 
the total $t$ dependent Hamiltonian $H$.  The use of $\tau_{1}(t)$ is natural within the frame of the adiabatic 
approximation \cite{Stockburger} where the system is assumed to be at thermal equilibrium at every instant. In our fully 
quantum approach, the question is much more difficult. In the particular case of the low frequency real part of the 
susceptibility, however, one can easily explain why using either 
$\tau_1$ or $\tau_1(t)$ lead to very similar results. Indeed, $\tau_1(t)$ and $\tau_1$ mainly differ only  
for the TLS's whose gap lie in the range $\epsilon \leq p_0E$. But, as it will be shown in the insets of Fig. 1 and Fig. 3, 
the gaps of the TLS's driven in the nonlinear regime by a given $E$ extend on a much larger domain (see sections {\bf I.B)} 
and {\bf  II.B)}): this is one of the main results of our fully quantum approach. Thus the possible time dependence of 
$\tau_1$ is not expected to change the results. This was carefully checked by performing all the calculations reported here 
twice, 
once using $\tau_1$, once using $\tau_1(t)$:  the resulting differences between both assumptions turned out in any case to 
be totally negligible. Hence, throughout the paper $\tau_1$ is considered as time independent, by simplicity. 

With the above relations, we get for the diagonal elements $<\rho_{1,1}(t)>$ and $<\rho_{2,2}(t)>$ :

$$<\rho_{1,1}\left(t\right)>=\frac{1}{2}+\frac{\Omega_z}{2\sqrt{\Omega_x^2+\Omega_z^2}}\tanh{\frac{\hbar 
\sqrt{\Omega_x^2+\Omega_z^2
}}{2k_BT}} , $$ 
and 

$$<\rho_{2,2}\left(t\right)>=\frac{1}{2}-\frac{\Omega_z}{2\sqrt{\Omega_x^2+\Omega_z^2}}\tanh{\frac{\hbar 
\sqrt{\Omega_x^2+\Omega_z^2
}}{2k_BT}} . $$

For its off-diagonal elements, it is found :

$$<\rho_{1,2}\left(t\right)>=<\rho_{2,1}\left(t\right)>=\frac{\Omega_x}{2\sqrt{\Omega_x^2+\Omega_z^2}}\tanh{\frac{\hbar 
\sqrt{\Omega
_x^2+\Omega_z^2}}{2k_BT}} , $$

where

 $$ \Omega_x\left(t\right)= - 2 \frac{\Delta_0}{\epsilon} \frac{\mathbf{p_0}.\mathbf{E}}{\hbar} \cos\omega t , $$

$$ \Omega_z\left(t\right)= - \frac{\epsilon}{\hbar} - 2 \frac{\Delta}{\epsilon} \frac{\mathbf{p_0}.\mathbf{E}}{\hbar} 
\cos\omega t . $$

Finally, one finds for the phonon field contribution:

$$ <\bar{S}_x>= \frac{\hbar\Omega_x}{2\sqrt{\Omega_x^2+\Omega_z^2}} \tanh{ \frac{\hbar 
\sqrt{\Omega_x^2+\Omega_z^2}}{2k_BT}} , 
$$

$$ <\bar{S}_z>= \frac{\hbar\Omega_z}{2\sqrt{\Omega_x^2+\Omega_z^2}} \tanh{ \frac{\hbar 
\sqrt{\Omega_x^2+\Omega_z^2}}{2k_BT}} , 
$$

and $<\bar{S}_y>=0$.

\vskip0.5cm

\textit{ii)} "Spin-spin" induced relaxation

Let us now turn to "spin-spin" interactions : for a given TLS, the effects of thermal transitions of its neighboring TLS's 
can 
be modeled as a small (fluctuating in time)
 electric field, i.e., as small fluctuating terms $\delta H(t) \ll \epsilon,k_{B}T$.
 The latter inequality ensures that the relaxation of the population of the levels (involving $\bar{S_{z}}$) will not be 
sensitive to $\delta H(t)$. 
It is shown in the Appendix {\bf A} that, for \textit{a given} TLS, the oscillations of $S_{x,y}(t)$  are no longer regular 
but progressively 
deformed by the random $\delta H(t)$ terms: due to the absence of correlations between the $\delta H(t)$ values seen by 
various TLS's, \textit{ensemble averaging} leads, by cancelation of phases of many TLS's \cite{Yu}, to a relaxation of 
$\bar{S}_{x,y}$ to \textit{zero} (while $S_{x,y}$ remains finite for any given TLS).
 This happens on a short characteristic time scale $\tau_{2} \ll 0.1\omega^{-1}$ and 
 yields a supplementary $\bar{S}_{x,y}/\tau_{2}$ for the relaxation of the coherence terms.

The temperature 
dependence of $\tau_{2}$ is not clear at present : in echo experiments \cite{Piche}, \cite{Bernard}, both $\tau_{2} \propto 
T^{-1}$  
as well as $\tau_{2} \propto T^{-2}$  were reported \cite{Graebner}. This might come both from the fact that accounting for 
the detailed shape of echo signals requires a very subtle theory (see e.g. \cite{Laikhtman}) and from the fact that several 
mechanisms contributes to $\tau_2$. Indeed, the pioneering work \cite{Halperin} of 
Black 
\textit{et al.} predicted a $\tau_{2} \propto T^{-2}$ dependence but very recent calculations \cite{Burint2} based upon the 
mechanism 
used in 
part {\bf{II}} found that $\tau_2 \propto T^{-1}$ could be justified at low $T$. Since this new mechanism will be used in 
the 
last section, we use throughout this work $\tau_{2} =\kappa_2/ T$, where $\kappa_2$ is a sample dependent constant. In 
order to try to take into account the various mechanisms which might contributes to $\tau_2$, the parameter $\kappa_2$ will 
be widely varied, as can be seen in Fig. 2. Last, owing to the smallness of the $p_0E$ values considered here, we neglect any 
$E$ effect on $\tau_2$ as explained in Appendix {\bf B}.
\vskip0.5cm

\textit{iii)} Final form of the Bloch equations

  Inserting the above relaxation terms in Eq.(2), the three Bloch equations can be written as follows:

$$ \frac{d\bar{S}_x}{dt} - \Omega_z\left(t\right) \bar{S}_y + \frac{\bar{S}_x-<\bar{S}_x>}{2\tau_1} + 
\frac{\bar{S}_x}{\tau_2} 
=0 , 
\eqno(3a) $$

$$ \frac{d\bar{S}_y}{dt} - \Omega_x\left(t\right) \bar{S}_z + \Omega_z\left(t\right) \bar{S}_x + \frac{\bar{S}_y}{2\tau_1}+
\frac{\bar{S}_y}{\tau_2} =0, \eqno(3b) $$

$$ \frac{d\bar{S}_z}{dt} + \Omega_x\left(t\right) \bar{S}_y + \frac{\bar{S}_z-<\bar{S}_z>}{\tau_1} =0 , \eqno(3c)$$

where all the $\bar{S}/\tau$ terms come from the relaxation processes, while all the $\Omega \bar{S}$ terms arise from the 
quantum 
dynamics, i.e. from the fact that $H$ and $\mathbf{s}$ do not commute.

Equations (3a) and (3b) also write

$$ \frac{d\bar{S}_+}{dt} + i\Omega_z\left(t\right) \bar{S}_+ + \frac{\bar{S}_+}{\tau_2^*} = i \Omega_x\left(t\right) 
\bar{S}_z 
+
\frac{<\bar{S}_x>}{2\tau_1}, \eqno(4) $$

with $$\bar{S}_+= \bar{S}_x+i\bar{S}_y,$$ and $\tau_2^*= \frac{2\tau_1\tau_2}{2\tau_1+\tau_2} . $

Let us note that $\tau_2^*$ appears due to the existence in eqs.(3a)-(3b) of the two terms $\bar{S}_{x,y}/(2\tau_1)$. Even 
if 
they are required by consistency (see above and Ref. \cite{2t1}), these two terms do not exist in the pionneering works 
accounting either for the 
small 
instationnarities  
\cite{Caruzzo} or for echo experiments \cite{Piche}, \cite{Bernard}, \cite{Graebner}. In fact these two terms play a 
\textit{negligible role} in the nonlinear 
susceptibility. To show this, let us first note that as long as $\tau_1 > \tau_2$, one gets $\tau_2^* \simeq \tau_2$, i.e. 
the Eqs. 
(3a)-(3c) amount to the simpler Bloch 
equations used before (especially in pulse echo experiments). The key point is that, in the $(\Delta, \Delta_0)$ plane, 
this domain 
where $\tau_1>\tau_2$ is \textit{quite large} : it is shown in the inset of Fig. 1 and in Ref. \cite{note12} that this 
domain 
contains, at least, all the TLS's such that $\epsilon \le e_{1,2} = 
(\kappa_{1}T/\kappa_2)^{1/3}$. As shown in the inset of Fig. 1, $e_{1,2} \simeq 0.2$ K is much larger than the $p_0E$ 
values studied 
in this work. This indicates that the TLS's standing \textit{out of} the $\tau_1 > \tau_2$ domain should not be affected by 
$E$, i.e. 
they should be in the linear regime (see Ref. \cite{exact}). To summarize, nonlinear effects should come mainly from the 
$\tau_1 > 
\tau_2$ region where the two terms $\bar{S}_{x,y}/(2\tau_1)$ are negligible. This  will be analytically demonstrated in 
section {\bf 
B)}\textit{2)}.

\subsubsection{Non perturbative resolution of the Bloch equations}

The Bloch equations cannot be solved analytically and even their numerical resolution is so far a great challenge.  
However, 
in the
audio-frequency range, some approximations can be made which strongly simplify the calculations.  As $\tau_2^{*}$ is much 
shorter 
than the
typical time $(\sim \frac{0.1}{\omega})$ to modify the populations, $\bar{S}_z$ may be considered constant 
\cite{AtomPhoton} 
in the right hand-side of Eq. (4).  The coherence terms follow adiabatically the population evolution.  They reach at every 
moment the stationary state
corresponding to the "frozen" occupation numbers.

Therefore, Eq. (4) can be solved independently of Eq. (3).  The stationary solution of Eq. (4) is

$$ \bar{S}_+= \frac{i\Omega_x\bar{S}_z + <\bar{S}_x>/{2\tau_1}}{i\Omega_z+1/\tau_2^*}, \eqno(5) $$

which inserted into Eq. (3) leads to a differential equation for $\bar{S}_z$:

$$ \frac{d\bar{S}_z}{dt} + \frac{\Omega_x^2/\tau_2^*}{\Omega_x^2+1/{\tau_2^*}^2}\bar{S}_z + 
\frac{\bar{S}_z-<\bar{S}_z>}{\tau_1} =
\frac{\Omega_x\Omega_z}{\Omega_x^2+1/{\tau_2^*}^2}\frac{<\bar{S}_x>}{2\tau_1}, \eqno(6) $$

$\bar{S}_z(t)$ in Eq.  (6) is expanded into its Fourier series to get its stationary state.  The expansion is limited to a 
finite number of harmonics.
This number, of the order of $10$, is found \textit{a posteriori} when a stable and accurate result is obtained.  So the 
differential equation is equivalent to a linear
system whose solutions are the harmonics $\bar{S}_{z}^{n}$.  The inverse Fourier-transform gives the periodic evolution of
$\bar{S}_z(t)$.  The coherence terms $\bar{S}_x$ and $\bar{S}_y$ are deduced from Eq. (5) 
where $\bar {S}_{z}(t)$, the solution of Eq. (6), is inserted. Finally, the 
first harmonics $\bar{S}_{x}^{1}$ of $\bar{S}_{x}(t)$ is sought, to be included into the dielectric susceptibility (see Eq. 
(7) below).

Indeed, the susceptibility \cite{Caruzzo} of a single TLS reads 

$$ \bar{\chi}= \frac{-2\left|\mathbf{p_0}\right|}{\left|\mathbf{E}\right|}\cos\theta \left(
\frac{\Delta}{\epsilon}\frac{\bar{S}_{z}^{1}}{\hbar} + \frac{\Delta_0}{\epsilon}\frac{\bar{S}_{x}^{1}}{\hbar} \right) , 
\eqno(7) $$

and it must be averaged over the distribution of TLS's \cite{Caruzzo} and over the dipole-orientation angle $\theta$ 
 to yield the total susceptibility of the sample : 

$$ \chi= \overline{P} \int^{\Delta_{max}}_0 d\Delta \int^{\Delta_{0max}}_{\Delta_{0min}} \frac{d\Delta_0}{\Delta_0} 
\int^1_{-1}
d\left(\cos\theta\right) \bar{\chi}\left(\Delta,\Delta_0,\theta\right) . \eqno(8)$$

 In the remainder of this article, we concentrate on the real
part $\chi'$ of $\chi$ which is linked to the capacitance of the sample, i.e., to its dielectric constant $\epsilon_r$ by : 
$$\epsilon_{r}-1 = \frac{\chi'}{\epsilon_0}.$$

\subsection{The quantum saturation effect: the quantum part of $\chi'(T)$ is depressed by a $E$ increase}

\subsubsection{Numerical results}

We have used the standard values for amorphous-$\mathrm{SiO_2}$:  $p_0=1$ $\mathrm{D}$, $\overline{P}= 3\times10^{44}$ 
$\mathrm{Jm^{-3}}$, $\kappa_{1}=10^{-8}$
  $\mathrm{sK^{3}}$ (all the energies in $\tau_1$ taken in K), $\Delta_{0min}= 10^{-6}$ K, $\Delta_{0max}= 10$ K, 
$\Delta_{max}= 10$ K. As explained above, we took $\tau_{2} =\kappa_{2}/T$, where $\kappa_{2}$ was ranged from $3.10^{-11}$ 
sK 
to 
$10^{-7}$ sK, allowing to check our fundamental assumption $\omega \tau_2 \ll 1$ provided $T\ge 0.5$ mK. Last, the 
numerical 
relative accuracy of our 
simulations was, in any case, better than $10^{-3}$ : this was checked very carefully, both by 
increasing the number of harmonics when solving Eq. (6) and by 
letting the successive integration procedures converge to better than $10^{-4}$. For each set of parameters $E,T, 
\kappa_{1}, 
\kappa_{2}, \Delta_{0,min}$ at least $4 \times 10^{4}$ couples of $(\Delta,\Delta_{0})$ were computed.

The simulations are displayed on Fig. 1.  The resonant response (low temperature) is strongly depressed by the drive level, 
while the
relaxation contribution (high temperature) is little affected.  This is at odds with the experiments \cite{Rogge} where 
increasing $E$ 
leads to an 
increase of both
the resonant response and of its slope $\left\vert \partial \epsilon'_r/\partial T \right\vert$ below $T_{rev}$. Let us 
note 
that the curve labeled 
"linear response" was obtained \textit{ independently} by a standard series expansion of the Bloch equations keeping only, 
as 
in Ref. \cite{Caruzzo}, the 
terms proportionnal to $E$ : as $E$ is made very small, the nonlinear calculations very precisely converge 
towards the linear regime.

\begin{figure} 
\includegraphics[height=6.5cm, width=8cm]{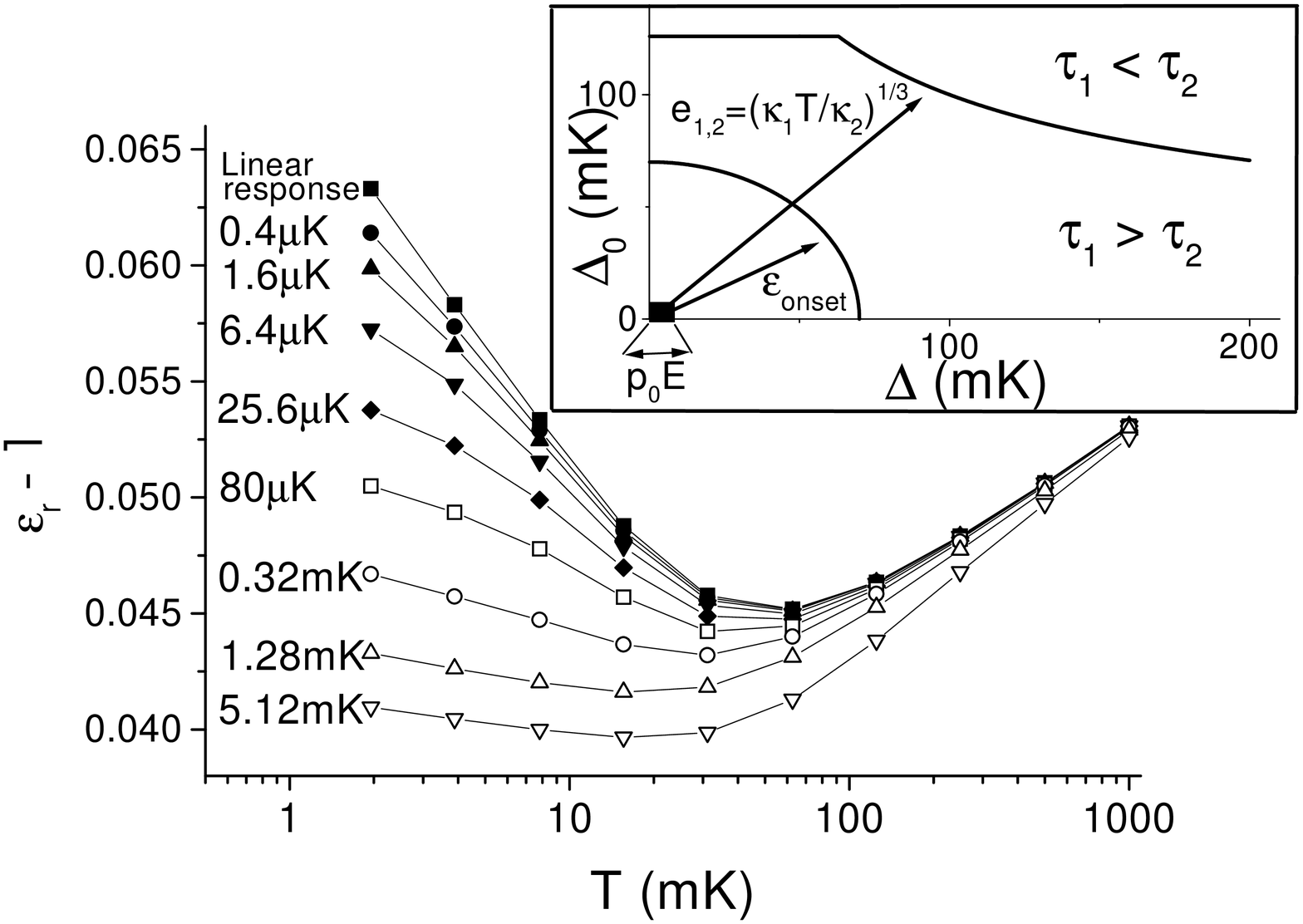} 
\caption{\textit{Inset:} At $T=10$ mK, $\kappa_{1}=10^{-8}$ $\mathrm{sK^{3}}$, 
 and $\kappa_{2}=10^{-8}$ $\mathrm{sK}$, the domain of TLS's such that $\tau_2<\tau_{1}$ is quite large and contains all 
the gaps 
smaller than $e_{1,2} = (\kappa_1T/\kappa_2)^{1/3}$ -see \cite{note12}-. Even at $p_0E =5.12mK$ this domain is larger than 
the one of the 
TLS's driven in 
the nonlinear regime defined by $\epsilon \le \epsilon_{onset} \simeq 70$ mK (see Eq. (11c) ). Note 
that $\epsilon_{onset} \gg p_0E$ ($p_0E$ is the small black area very 
near the origin): this explains that the nonlinear effects are visible even at very low fields, as shown in the main 
Figure. 
\textit{Main Figure: } Dielectric susceptibility of amorphous-$\mathrm{SiO_2}$ 
at 1 kHz vs temperature simulated at various fields  -the value of $p_0E$ in Kelvin labels each curve- within the standard 
two-level
 system model with the following set of parameters:
 $p_0=1$ $\mathrm{D}$,   
$\kappa_{1}=10^{-8}$ $\mathrm{sK^{3}}$, 
 $\kappa_{2}=10^{-9}$ $\mathrm{sK}$, 
$\Delta_{0,min}=10^{-6}$ $\mathrm{K}$,
  $\Delta_{max}=\Delta_{0 max}=10$ $\mathrm{K}$,
 $\overline{P}= 3\times10^{44}$ $\mathrm{Jm^{-3}}$. 
 The low-temperature response vanishes rapidly as the electric field is increased due to the quantum saturation phenomenon. 
The linear response was obtained by an independent perturbative method.}  
\label{Fig.1} 
\end{figure}

However, the extreme sensitiveness of the resonance to the external field is very striking.  It decreases rapidly while
$\left|\mathbf{p_0}.\mathbf{E}\right|<<k_BT$.  The low-temperature phase-coherent upturn is destroyed by its environment 
(the 
external
field), although the perturbation is much smaller than any thermodynamical quantity, which suggests that this effect has a 
quantum
origin. This is further confirmed by the inset of Fig. 2 showing the 
influence of $T$ and $\tau_2$ on $\delta \chi'(E,T) = 1-\chi'(E,T)/\chi'(0,T)$ : for a given $E$, the smaller $T$, the 
larger 
$\delta \chi'$, which is expected since quantum effects generally increase as $T$ decreases. Similarly, $\delta \chi'$ is 
larger when $\kappa_{2}$ is made smaller, i.e., when quantum coherence is made more "fragile". Finally, the dimensionless 
$\delta \chi'$ appears to depend not only on $E,T, \kappa_{2}$ but also on $\kappa_{1}$, and it is shown in the main part 
of 
Fig. 2 that all these dependencies are a \textit{universal} function of a dimensionless scale $\eta$, namely, : 

$$\delta \chi'=
\left\{
\begin{array}{ccc}
0.1 \times \sqrt{\eta} & if & \eta \lesssim 1 \\
0.1 \times \ln(\eta)& if & \eta \gg 1 \\
\end{array}
\right.
\hbox{ with } \eta={\frac{p_0E}{k_BT}}(\frac{ \kappa_{1}}{T^2 \kappa_{2}})^{\alpha} ,  \eqno(9)$$

where $\alpha\simeq 0.45 \pm .05$ and $\ln(\eta)$ might be replaced by a power law of $\eta$ with an exponent lower than 
$0.1$. This universal $\delta \chi'(\eta)$ dependence holds only when the relaxational part of $\chi'$ can be totally 
neglected, i.e., well below $T_{rev} \simeq 50$ mK : in Fig. 2, only data corresponding to $T\le 10$ mK have been plotted. 
For 
these low $T$, $\delta \chi'(\eta)$ remains universal even when $(\kappa_{1}, \kappa_{2}, E)$ are varied over several 
decades. 
The factor $\kappa_{1}/(T^2\kappa_2)$ in $\eta$ becomes very large at low $T$, yielding nonlinear effects even for very 
small 
$E$: this expresses that the lower $T$, the smaller the onset field of the nonlinear 
regime, as already seen on Fig. 1. Let us mention that the data of Fig. 2 correspond to the 
particular case $\theta =0$.

\begin{figure}
 \includegraphics[height=7.5cm, width=8cm]{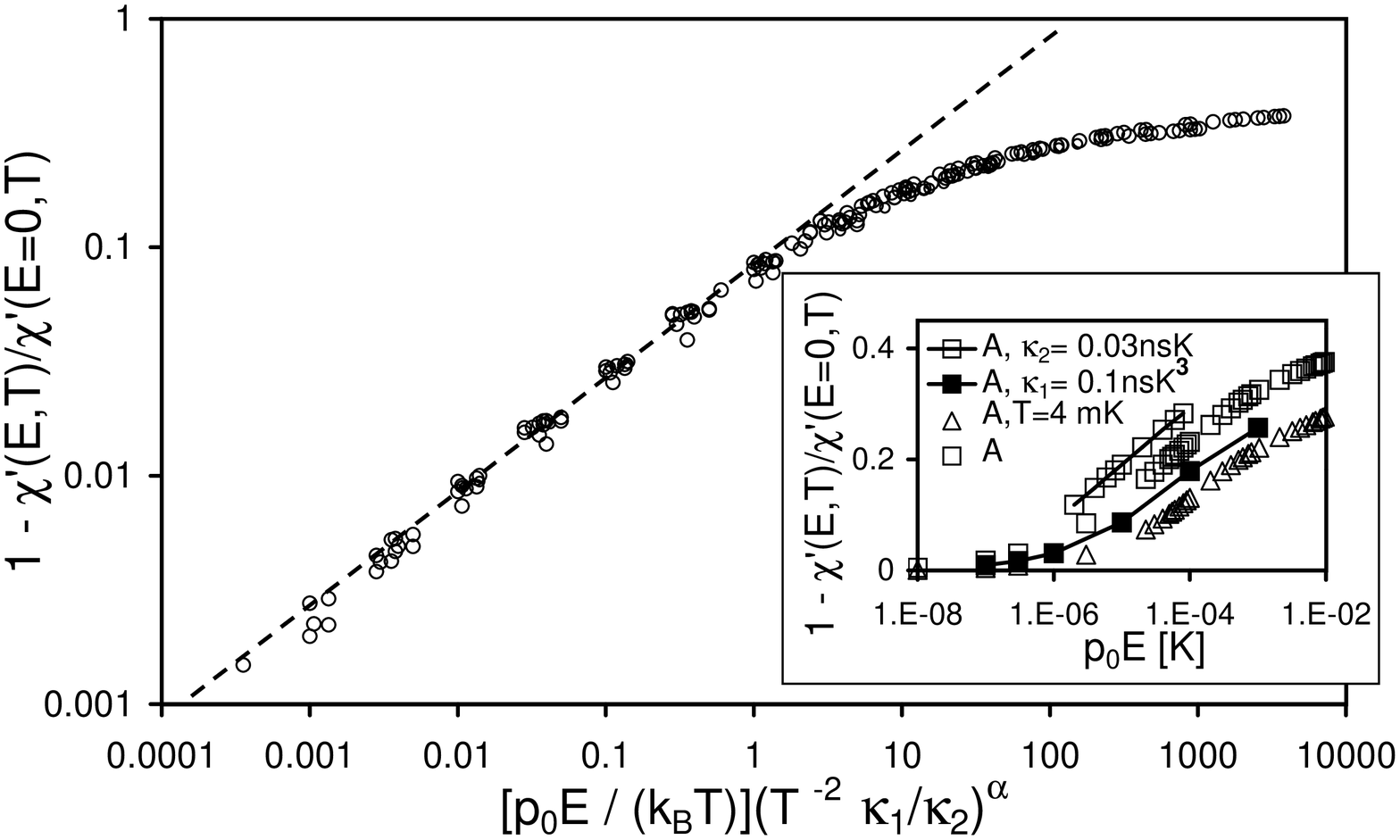}
 \caption{\textit{Inset :} $\delta \chi' = 1-\chi'(E,T)/\chi'(0,T)$ 
plotted versus $p_0E$ (in kelvins). Curve A corresponds to 
$p_0=1$ $\mathrm{D}$,   $\kappa_{1}=10^{-8}$ $\mathrm{sK^{3}}$,  $\kappa_{2}=10^{-9}$  $\mathrm{sK}$, 
$\Delta_{0,min}=10^{-6}$ $\mathrm{K}$,  
$\Delta_{max}=\Delta_{0 max}=10$ $\mathrm{K}$, 
$\overline{P}= 3\times10^{44}$ $\mathrm{Jm^{-3}}$ 
and $T=2$ $\mathrm{mK}$. 
The other three curves show the effect upon quantum saturation of the
 parameter which was changed with respect to A : increasing $T$,
 as well as decreasing $\kappa_{1}$, decreases $\delta \chi'$; while
 decreasing $\kappa_{2}$ increases $\delta \chi'$, as expected due
 to the quantum nature of $\delta \chi'$. \textit{Main figure} :
 The various influences of the simulation parameters can be reduced 
to a universal function of the dimensionless variable $\eta={\frac{p_0E}{k_BT}}(\frac{ \kappa_{1}}{T^2 
\kappa_{2}})^{\alpha}$
 with $\alpha = 0.45 \pm .05$ numerically. The dashed line shows that 
$\delta \chi' \propto \sqrt{\eta}$ when $\eta \lesssim 1$. The various
 parameters were ranged over several decades : 
$10^{-10}$ $\mathrm{sK^{3}}$$\le \kappa_{1} \le 10^{-8}$ $\mathrm{sK^{3}}$;
 $3 \times10^{-11}$ $\mathrm{sK}$$\le \kappa_{2} \le 10^{-7}$ $\mathrm{sK}$;
  $10^{-6}$ $\mathrm{K}$$\le\Delta_{0,min} \le 10^{-4}$ $\mathrm{K}$; 
$10^{-8}$ $\mathrm{K}$$\le p_0E \le 3$ $\mathrm{mK}$. The data of this figure correspond to the particular case $\theta 
=0$.}
 \label{Fig.2}
 \end{figure}

\subsubsection{Physical interpretation}
To further understand the universal $\delta \chi'(\eta)$ and demonstrate its quantum origin, let us briefly go into the 
structure of the Bloch equations. By 
using the identity $\Omega_{x}<\bar{S}_{z}>=\Omega_{z}<\bar{S}_{x}>$, Eq. (6) can be written : 

$$\frac{d\bar{S}_z}{dt} + \frac{\bar{S}_z}{\tau_{z}} = \frac{<\bar{S}_z>}{\tau_{z,1}}
\hbox{,} 
\left\{
\begin{array}{ccc}
\frac{1}{\tau_{z,1}} &=& \frac{1}{\tau_{1}}\left(1+ 
\frac{1}{2}\frac{(\Omega_{x}\tau_{2}^{*})^2}{1+(\Omega_{z}\tau_{2}^{*})^2} 
\right) \\
\frac{1}{\tau_{z}} &=& \frac{1}{\tau_{1}}\left(1+ 
\frac{\tau_1}{\tau_2^{*}}\frac{(\Omega_{x}\tau_{2}^{*})^{2}}{1+(\Omega_{z}\tau_{2}^{*})^{2}} \right) \\
\end{array}
\right.  , \eqno(10)$$

In Eq. (10),  one gets at $E \to 0$ : $\tau_z = \tau_{z,1}=\tau_1$. As argued in section {\bf I)A)}\textit{2)}, the 
nonlinear 
behavior should come from the TLS's such that $\tau_{1}>\tau_2$: in this case we see indeed from Eq. (10) that increasing 
$E$ 
decreases $\tau_{z}$ \textit{much more} than 
$\tau_{z,1}$. This strongly depresses the off diagonal 
suceptibility, as we show now. 

Let us first derive, from Eq. (10), the critical value $E^*$ such that  $1/\tau_{z}$ becomes larger than $1/\tau_{z,1}$: 
focusing on the 
gaps $\epsilon$ lying within the $\tau_1> \tau_2$ domain, i.e. in the domain where $\tau_2^* \simeq \tau_2$, $E^*$ 
is determined by the condition  $\tau_{1} \tau_{2} \Omega_{x}^{2} \simeq 1+\Omega_{z}^{2}\tau_{2}^{2}$, yielding :

$$\left\{
\begin{array}{ccccr}
\frac{p_0E^*}{\epsilon}&=& \frac{\hbar T}{k_B \sqrt{\kappa_{1}\kappa_{2}}}& \hbox{ if }k_B\epsilon\tau_2 \le \hbar& (11a)\\ 
\frac{p_0E^*}{\epsilon}&=& \epsilon\sqrt{\frac{\kappa_{2}}{\kappa_{1}}}& \hbox{ if }k_B\epsilon\tau_2 \ge \hbar& (11b)\\
\end{array}
\right. , 
$$

where all the energies are expressed in kelvins. With the standard values $\kappa_{1} = 10^{-8}$ $\mathrm{sK^3}$ and 
$\kappa_{2}= 10^{-9}$ $\mathrm{sK}$, we see that $p_0E^*$ is \textit{much smaller} than $\epsilon$. Indeed, for $T=10$ mK 
we get  
$p_0E^{*}/\epsilon = 2.10^{-5}$ for the smallest 
gaps following Eq. (11a), and, for example, $p_0E^{*}/\epsilon  \le 3.10^{-3}$ for the gaps $\epsilon \simeq k_BT$ which 
follow Eq. (11b). Solving  Eq. (11b) with respect to $\epsilon$, for \textit{a given} $E$, leads to a characteristic gap 

$$\epsilon_{onset} = \sqrt{p_0E} \left( \frac{\kappa_1}{\kappa_2} \right)^{1/4} , \eqno(11c)$$

where all the energies are in Kelvins. For the highest $p_0E \simeq 5.12$ mK studied here, we get $\epsilon_{onset} \simeq 
70$ mK. As 
shown in the inset of Fig. 1, $\epsilon_{onset}$ is both much larger than $p_0E$ and corresponds to a domain smaller than 
the one 
defined by our assumption $\tau_1>\tau_2$. 

To show that $E^*$ in Eq. (11b) is indeed the critical field for a given TLS, at which the kind of nonlinearities of Figs. 
1-2 
onsets, let us 
now compare $\chi'(E \ll E^*)$ and $\chi'(E^*)$.
 
\textit{i) If }$E \ll E^{*}$, we get from Eq. (10) $\tau_{z} \simeq \tau_{z,1} \simeq \tau_{1}$. Solving Eq. (10) is 
straightforward and leads for the $n^{th}$ harmonics of 
$\bar{S}_{z}(t)$ :

$$\bar{S}_{z}^{n} = \frac{<\bar{S}_{z}^{n}>}{1+n^{2}\omega^2 \tau_{1}^{2}} , 
\eqno(12)$$

where $<\bar{S}_{z}^{n}>$ is the $n^{th}$ harmonics of $<\bar{S}_{z}(t)>$. Remembering that the region of interest is 
$\epsilon < 
\epsilon_{onset}$, 
it can be checked that $\omega \tau_{1} \gg 1$ for basically all the considered 
TLS's. This yields, from Eq. (12), $\bar{S}_{z}(t) \simeq <\bar{S}_{z}^{0}>$. Furthermore, since 
$p_0E \ll \epsilon$ due to Eqs. (11), we get 
$<\bar{S}_{z}(t)> \simeq <\bar{S}_{z}^{0}>$, which, once combined with the identity $\Omega_{x} <\bar{S}_{z}(t)> = 
\Omega_{z} 
<\bar{S}_{x}(t)>$,  
yields $\bar{S}_{z}(t) \simeq \Omega_{z}<\bar{S}_{x}(t)>/\Omega_{x}$. Once reported into Eq. (5), this yields :

$$\bar{S}_{x}(t) \simeq \frac{<\bar{S}_{x}(t)>}{1+\Omega_{z}^{2} \tau_{2}^{2}}
(\Omega_{z}^{2} \tau_{2}^{2})  , \eqno(13)$$

where in the last factor the fact that $\Omega_{z}^{2}\tau_{2}^{2} \gg \tau_{2}/(2 \tau_{1})$, which holds for any 
reasonable 
set of $(\kappa_{1}, \kappa_{2})$, was used to drop the term $\tau_{2}/(2 \tau_{1})$.

\textit{ii) For }$E = E^{*}$, we get from Eq. (10), $\tau_{z,1} \simeq \tau_{1}$ and $\tau_{1}/2 \le \tau_{z}(t) \le 
\tau_{1}$. The fact that $\tau_{z}$ is now smaller than $\tau_{z,1}$ is responsible for the onset of nonlinear 
effects. This can be seen by setting $\tau_{z} = \tau_{1}/2$ throughout the electrical period. With this simplification, 
one 
gets, with a derivation similar to the one yielding Eq. (13) :

$$\bar{S}_{x}(t) \simeq \frac{<\bar{S}_{x}(t)>}{1+\Omega_{z}^{2} \tau_{2}^{2}}
(\frac{1}{2}\Omega_{z}^{2} \tau_{2}^{2}) , \eqno(14)$$

The off-diagonal part of the response in phase with $E$ is $\bar{\chi}'_{x}\propto \bar{S}_{x}^{1}/E$ : it is read directly 
from Eqs. (13)-(14), remembering that $<\bar{S}_{x}> \propto E\cos\omega t$. This yields $\bar{\chi}'_{x}(E = E^*) \simeq 
\frac{1}{2} \bar{\chi}'_{x}(E \ll E^*)$, where the factor $1/2$ comes from the above relation $\tau_{z} = \frac{1}{2} 
\tau_{z,1}$, which was a simplification of the case $E=E^*$. The comparison of Eqs. (13)-(14) is thus only 
semi-quantitative,  
but it 
yields the main two features of the quantum saturation phenomenon : first $\bar{\chi}'(E^*) < \bar{\chi}'(E \ll E^*)$, 
second 
this effect comes from the off-diagonal part of the susceptibility, i.e., it is purely quantum (the diagonal susceptibility 
$\bar{\chi}'_{z} \propto \bar{S}_{z}^{1}/E$ is much smaller than $\bar{\chi}'_{x}$ due to the fact that $\omega \tau_{1} 
\gg 
1$ below $T_{rev}$).

We have here an example of quantum decoherence \cite{Fisher}. It is not surprising that these effects were missed by 
the adiabatic approximation mentionned in the introduction since, in this approach, $\tau_2$ has disappeared, yielding for 
the 
nonlinear onset \cite{Rogge} no other possibility than 
$\left|\mathbf{p_0}.\mathbf{E}\right| \sim k_BT$, as expected for a system at equilibrium. Moreover we have 
shown that the quantum saturation depends on the precise coupling of the three Bloch equations, i.e. of the fact that 
$\tau_{z}$ evolves faster with $E$ than $\tau_{z,1}$ : this is out of reach for the adiabatic approximation since it 
contains 
only \textit{one} differential equation \cite{Stockburger} instead of Eqs. (3a)-(3c). Finally, the results of Figs 1-2 do 
not depend on the precise microscopic mechanism involved in $\tau_2$, but only on the fact, well established by echo 
experiments, that, for a vast subclass of TLS's one has $\tau_2 \ll \tau_1$ : this is the main reason of the $E$-induced 
depression of $\chi'$ of Figs 1-2. 

\subsubsection{Effect of the density of states}

More can be learnt from Eqs. (11), and more precisely from Eq. (11b) which holds for the vast majority of the TLS's 
responsible for 
the nonlinear behavior. First, let us note that the onset field $E^*$ increases as $\sqrt{\kappa_{2}/\kappa_{1}}$ : this 
suggests 
that the depression of $\chi'$, when $E$ is increased, depends on $E\sqrt{\kappa_{1}/\kappa_{2}}$, which, remembering that 
$\kappa_{1}/\kappa_{2}$ is the square of a temperature, leads 
to the dimensionless scale $p_0E/(k_{B}T)\sqrt{\kappa_{1}/(T^{2}\kappa_{2})}$ as the natural
 parameter for the quantum saturation phenomenon. This dimensionless scale matches exactly  the definition of $\eta$ in Eq. 
(9).

 Second, from the above discussion of Eqs. (13)-(14), the TLS's such that $\epsilon \le \epsilon_{onset}$ are 
already in the saturation regime, while the gaps larger than $\epsilon_{onset}$ are hardly altered by $E$. 
It is thus 
natural to consider the number of TLS's such that  
$\epsilon \le \epsilon_{onset}$ as an estimate of the amplitude of the quantum saturation phenomenon 
$1-\chi'(E,T)/\chi'(0,T)$, stating : 

$$1-\chi'(E,T)/\chi'(0,T) \propto \int_{\epsilon_{min}}^{\epsilon_{onset}} P(\epsilon) d\epsilon \propto 
\sqrt{E} \propto \sqrt{\eta} , \eqno(15)$$

where the last equality was obtained by using the above-stated relationship $E \propto \eta$; while the second equality 
uses 
 both Eq. (11c) and the fact that the energetic density of 
states $P(\epsilon)$ is a constant due to the standard distribution
 $P(\Delta, \Delta_{0}) = \bar{P}/\Delta_{0}$. Equation (15) yields exactly Eq. (9) derived from  the numerical 
simulations. 
This argument enables to state that the small
 corrections to the standard $\bar{P}/\Delta_{0}$ previously proposed only yield small changes to the behavior of Figs. 1-2 
: 
this is true, e.g., for the slight depression of the density of states at small gaps derived by Burin \cite{Burin95}, as 
well 
as for $\bar{P}/\Delta_{0}^{1+y}$ with $\vert y \vert \ll 1$ proposed in Ref. \cite{Frossati77}.

To summarize this section {\bf I)}, solving the Bloch equations leads to the quantum saturation effect, i.e., 
to a strong decrease of the off-diagonal part of $\chi'$ when $E$ is raised. This effect holds for a very large set of 
$\kappa_1$ and $\kappa_2$ -the main parameters of the model-, and it mainly comes from the TLS's such that $\epsilon \le 
\epsilon_{onset} <  
e_{1,2}$. For an ensemble of TLS's with a $\bar{P}/\Delta_{0}$ density of states, quantum 
saturation goes 
as $E^{0.5}$, and such an exponent justifies \textit{ a posteriori} the 
nonperturbative character of the method used  here. Last, the quantum saturation phenomenon onsets for fields $E^* \ll k_B 
T/p_0$, as seen from Eq. (9). It is thus non-negligible
 since the field is, in most experiments, decreased well below $k_B T/p_0$. However, in the literature, the trend of the 
data 
is \textit{systematically the opposite} of the one of Figs. 1-2.   
Since -see Appendix {\bf B}- more general Bloch equations, corresponding to larger $E$,  should not qualitatively change 
the results of Figs. 1-2, we conclude that  the standard TLS model cannot account for the basic features of the nonlinear 
experimental data in the kHz range.

\section{Adding interactions}

\subsection{ Burin \textit{et al}'s mechanism}

At this step, at least one drive-dependent parameter must be added into the model to explain the large discrepancy with the 
experimental data. Moreover, it must
enhance the relaxation process at low temperature, since coherence is broken by the external field as shown in Figs. 1-2.

Recently, Burin \textit{et al.}  \cite{Burin} proposed an additional field-induced relaxation mechanism.  They show that 
the 
resonant
dipole-dipole coupling, which is so small in glasses, can be strongly increased by a low-frequency electric field.  Indeed, 
thermal excitations, which are at zero-field localized on each TLS, tend to delocalize by hopping to resonant nearest 
neighbors. This is due to the fact that \textit{resonant hopping} demands \textit{both} TLS's to have very close values of 
\textit{both} $\Delta$ \textit{and}  
$\Delta_{0}$ : as the electrical field modulates the TLS parameter $\Delta$, the probability of finding, for a given TLS, a 
resonant TLS,  increases from a negligible value at very low $E$, to a non-negligible value above a threshold of the 
external 
field. Let us note that this mechanism transports energy; hence it can be treated as a new
relaxation mode.

The frequency must be small for the electric field to have time to modulate the coupling parameters. This is of no 
consequence 
here, since our crucial assumption $\omega \tau_2 \ll 1$, leading to Eq. (5), already restricts our work to the low 
frequency 
case. Another assumption is 
that the external field amplitude is smaller than the
characteristic splitting energy $\sim k_BT$, in order to treat the field as a weak perturbation.  The typical values of the 
frequency and
$\left|\mathbf{p_0}.\mathbf{E}\right|$ are respectively 100 Hz and 1 mK but may be softened as a rigourous determination is 
out of reach.

When the electric field increases, so does the probability of finding a resonant neighbor close enough to yield tunneling 
with 
not too small a probability : the one-particle excitation will relax more
rapidly at high $E$ towards another site.  One can show the relaxation rate is proportionnal to the square root of the 
drive 
level \cite{Burin}. To include this new enregy relaxation channel, we set in Eqs.(3a)-(3c) $\tau_{1}^{-1} = \tau_{1, 
ph.}^{-1} 
+ \tau_{1B}^{-1}$ where $\tau_{1,ph}$ is the phonon field induced relaxation mechanism used throughout section {\bf I)} and 
where 

$$\tau_{1B} =
\frac{\cal B}{\sqrt{\left|\mathbf{p_0}.\mathbf{E}\right|}} , \eqno(16)$$

 with the constant ${\cal B} = 10^{-5}$ sK$^{1/2}$ for physically reasonnable parameters \cite{Burin}. As a result, 
increasing 
$E$ at any given $T$ leads to an \textit{increase} of the susceptibility 
$\chi'$ : this shows that Burin \textit{et al.}'s mechanism is strong enough to overcome the decrease due to the "quantum 
saturation phenomenon". However, the agreement between the set of calculated curves (unreported) and the data is very poor 
since the net increase of $\chi'(T)$ when $E$ is increased is \textit{stronger} at high $T$ than at low $T$. This is due to 
the fact that, since relaxation dominates the total response, the most influent TLS's are such that $\epsilon \le k_BT$ : 
their number enlarges with $T$ and so does their supplementary relaxational response due to the new relaxation channel 
$\tau_{1B}$. 

To interpolate between Fig. 1 and Eq. (16) which appear as extreme cases, one might use the very general argument that 
interaction effects should disappear at high $T$, e.g., above 
 100 $\mathrm{mK}$ -see Ref. \cite{Neu}-. This demands that the chosen $\tau_{1B}(T)$ becomes infinite, i.e., negligible, 
at high $T$. Such a 
general requirement can be of course modeled by different laws but all the ones we tried gave the same kind of behavior for 
the susceptibility. This is why we report on the calculations using a simple law, namely, 

$$\tau_{1B}\left(T\right)= \frac{\tau_{1B}}{1-e^{-T_B/T}} \ \ \hbox{with}\ T_{B}=15 \ \hbox{mK}, \eqno(17)$$ 

where $\tau_{1B}$ is given by Eq. (16) and the thermally activated
behavior models a dipole-dipole coupling constant of $T_B=15$ mK : the energy scale $T_B$ can be deduced from Fig. 3 of 
Rogge 
\textit{et 
al.}'s data \cite{Rogge} on $a$-$\mathrm{SiO_x}$  since $\chi'$ becomes $T$-independent below $15$ mK even for $E$ values 
ten 
times larger than the range of the linear regime. Of course, this $T_B$ scale can be adjusted empirically since the $T$ 
where 
$\chi'$ becomes $T$-independent depends on the material. As the coupling constant goes as 
$g/\left|\mathbf{r}-\mathbf{r'}\right|^3$
and as \cite{Caruzzo}, for $a$-$\mathrm{SiO_2}$, $g\sim10$ $\mathrm{Knm^3}$, we get a mean distance $\lambda_B$ between 
interacting 
dipoles of nearly $10$ nm.

\subsection{Numerical results}

The modified-model predictions using Eq. (17) are displayed on Fig. 3.  The values of 
$\left|\mathbf{p_0}.\mathbf{E}\right|$ 
have been limited to 10 mK
because of the restrictions on both the Bloch equations and the field-induced mechanism.  A trend  completely different 
from 
the one of Fig. 1 is obtained at low
temperature since an increase of 
the response is observed when the drive level increases.

\begin{figure} 
\includegraphics[height=6.5cm, width=8cm]{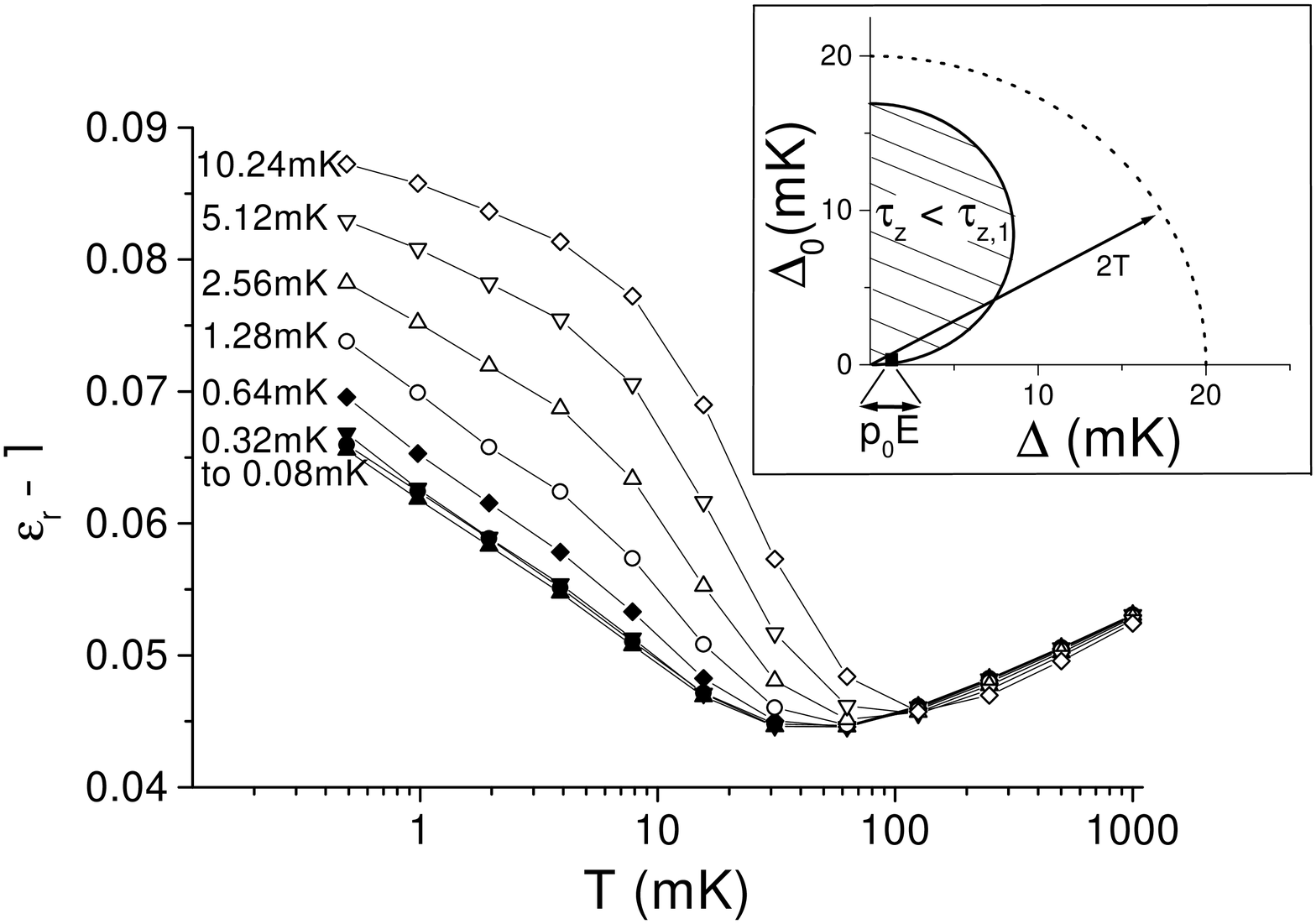} 
\caption{\textit{Main Figure : }Simulation of $a$-$\mathrm{SiO_2}$ susceptibility at 1 kHz vs temperature with Eq. (8) and 
the same parameters as in the main part of Fig. 1.  The calculations were done within a modified TLS model where 
excitations are no 
longer 
localized but can experience field-induced hops to neighboring sites, which is modeled by an additive relaxation channel 
(see 
the definition of $\tau_{1B}(T)$ in Eq. (17)).  The data show a linear behavior at low enough drive levels (the $p_0E$ 
values label the 
curves), an evolution of 
$T_{rev}$ with $E$ compatible with experiments and a substantial decrease of the $T$ dependence of $\chi'$ at low $T$ ;  
\textit{Inset :} 
For $p_0E \ll k_BT$, in the $(\Delta, \Delta_0)$ plane, $<\bar{S}_z^1>$ is not negligible only within the $\epsilon < 2T$ 
domain. Even 
for $p_0E = 0.8$ mK, the hatched area where $\tau_z < \tau_{z,1}$ has a non negligible size with respect to this $\epsilon 
< 2T$ domain : 
this yields a supplementary $T$-dependent contribution to the diagonal susceptibility $\chi'_z$ which overcomes the 
$E$-induced 
depression of $\chi'_x$ seen on Fig. 1, and yields the $E$-enhanced $\chi'$ trend seen on the main part of the figure.}  
\label{Fig. 3} 
\end{figure}

By computing separately (unreported) in Eq. (8) the two terms of the right hand side of Eq. (7), we checked that $\chi'_x$ 
behaves 
qualitatively as in section {\bf I)} and that the new trend of Fig.3 is  
due to the diagonal part $\chi'_z$. To explain this new behavior, one first note that $\tau_{1B}(T)$ is now the upper bound 
of $\tau_1$, 
even for the numerous TLS's whose small $\Delta_0$ value lead, in section {\bf I)}, to a very large $\tau_1$. With $\omega 
\tau_{1B}(T\lesssim T_B) \lesssim 1$, the $1/(\omega^2 \tau_1^2)$ cutoff of $\bar{S}_z$ seen on Eq. (12) has now 
disappeared, i.e. the 
$d\bar{S}_z/dt$ term in Eq. (6) can be dropped, yielding : 

$$\bar{S}_{z}(t) \simeq \frac{\tau_z}{\tau_{z,1}} \left<\bar{S}_{z}(t) \right> ,\eqno(18)$$

where $\tau_{z},\tau_{z,1}$ are defined in Eq. (10). At $E \to 0$, one has $\tau_z \simeq \tau_{z,1} \simeq \tau_1$, 
yielding with Eq. 
(18), $\bar{S}_{z}(t) \simeq \left<\bar{S}_{z}(t) \right>$. With the additionnal remark that 
$\left<\bar{S}_{z}^{1}(\epsilon<2T) \right> 
\simeq \hbar p_0E/(4k_BT)$ while $\left<\bar{S}_{z}^{1}(\epsilon>2T) \right> \simeq 0$, one gets, with the standard 
$\bar{P}/\Delta_0$ 
density of states, that $\chi'_z(T) \propto + \ln{T}$ : this is the trend seen above $T_{rev}$. 

To explain the behavior below $T_{rev}$, the key point is that for quite a large domain in the $(\Delta, \Delta_0)$ one has 
$\tau_z/ 
\tau_{z,1} < 1 $ : since this factor is $T$ dependent, it will modify the $T$ dependence just above derived for $\chi'_z$ 
from Eq. (18). 
Focusing on the $\epsilon \le 2T$ gaps, we get, below $T_{rev}$, $\tau_{1}\simeq \tau_{1,B}(T)$ and $\tau_1 \gg \tau_2$ : 
the condition 
$\tau_z/ \tau_{z,1} < 1 $ amounts to $\tau_{1} \tau_{2} \Omega_{x}^{2} \ge \Omega_{z}^{2}\tau_{2}^{2}$, i.e. :

$$\epsilon \le 2 p_0E \left( \sqrt{\tau_{1B}(T)/\tau_2} \ \sin{\phi} + \cos{\phi} \right)\ \hbox{with}\  \phi = 
\arctan{\frac{\Delta_0}{\Delta}} , \eqno(19)$$

The $\tau_z/ \tau_{z,1} < 1 $ condition is shown, as a hatched domain, in the inset of Fig. 3. Even for the lowest $E$ 
studied here, it 
is not negligible with respect to the $\epsilon < 2T$ area. Since in the hatched domain one has $\tau_{z}/\tau_{z,1} \simeq 
\tau_2 
\Omega_z^2/(\tau_1 \Omega_x^2)$, this factor remains $T$ dependent even below $T_B$ when $\tau_{1B}(T)$ has reached its 
maximum value : this is due to the fact that $\tau_2$ \textit{remains $T$ dependent even at very low} $T$. 
With $\left<\bar{S}_{z}^{1}(\epsilon<2T) \right> \simeq \hbar p_0E/(4k_BT)$, integration of Eq. (18) within the hatched 
area yields a 
contribution $\delta\chi'_z \propto E^{3/4}/T^{1/2}$. Thus: \textit{ i)} this term increases as $T$ decreases; \textit{ii)} 
$\delta\chi'_z$ increases with $E$, i.e. it can overcome the  $E$-induced depression of $\chi'_x$. Disregarding the slight 
difference 
-see \cite{notefin}- between the $\delta \chi'_{x} \propto -E^{1/2}$ seen for the quantum saturation phenomenon and the 
$\delta\chi'_z 
\propto +E^{3/4}$, the linear regime of Fig. 3, up to $p_0E=0.32mK$ can be seen as resulting from the compensation of both 
effects. At 
higher $E$, the $\delta\chi'_z$ increase dominates over the $E$-induced depression of $\chi'_x$, yielding a net increase of 
$\chi'$ with 
$E$. Note that $\chi'_z(E)$ becomes $T$ 
independent when $T \le p_0E/k_B$ : in this case, indeed, $\left<\bar{S}_{z}^{1}(\epsilon<2T) \right>$ is no longer $T$ 
dependent. This 
yields the substantial decrease of the $T$ dependence of $\chi'$ seen for the two highest $E$ values on Fig. 3.

Last, the off-diagonal susceptibility $\chi'_x \propto \bar{S}_x^1$ mainly behaves as in section {\bf I)}, i.e. we recover 
the 
quantum saturation phenomenon yielding, when $E$ is raised, both a decrease of $\chi'_x$ and of the slope $\vert \partial 
\chi'_x 
/\partial T \vert$. With respect to section {\bf I)} the quantum saturation effect is somehow weakened, which can be 
understood since, 
for a given $E$, the number of TLS's lying within the $\epsilon \le \epsilon_{onset}$ domain of Fig. 1 is larger than the 
corresponding 
one in the hatched area of Fig. 3. Finally, the variations of 
$\chi'_x$ with $T$ remain smaller than the ones of $\chi'_z$, excepted in the case where  
$T < p_0E/k_B$ : the small $T$ dependence of $\chi'(T<T_B; p_0E \gtrsim 5$ mK$)$ is thus the only case where $\chi'_x$ 
dominates the $T$ 
behavior of $\chi'$ on Fig. 3.

To summarize, the bigger the electric field, the smaller the field-induced relaxation time (see Eqs. (16)-(17)), which 
enhances the 
relaxationnal part of the response, leading to a net increase of $\chi'$ with $E$ at a given $T$. At a given $E$, when $T$ 
decreases 
below $T_B$, the $\chi'(T)$ increase is due to the fact that $\tau_2$ is still $T$ dependent : this is, of course, out of 
reach for the 
adiabatic approximation where $\tau_2$ has disappeared. Finally, inserting Burin \textit{et al.}'s new relaxation rate in 
Bloch equations 
allow to account for the main trend of the nonlinear data : however, in this approach, the so-called 
"resonant" regime below $T_{rev}$ is not a coherent
one but, mainly, a field-enhanced relaxation regime.

\subsection{Comparison with experiments}

On Fig. 3, one observes a pseudolinear regime up to $p_0E \simeq 0.05k_BT$ where the dielectric response is 
quasi-independent 
on the external field.  This value of
the electrical field agrees with the experimental linear regime, which, depending on the materials, extends up to 
$p_0E/(k_BT)$ 
in the range $[0.02; 0.12]$ (see Figs. 3-5 of Ref. \cite{Rogge}). We checked that this pseudolinear regime comes from the 
form 
of 
$\tau_{1,B} \propto E^{-\beta}$ where $\beta$ takes the highly nontrivial value $1/2$. Setting lower values for $\beta$, 
such 
as $\beta=0.1$, yields the quantum saturation phenomenon to dominate, leading to the same trends as in Fig. 1, at odds with 
experiments. Setting $\beta = 1$ leads to the tendency of Fig. 3 but with a linear regime reduced to  $p_0E/(k_BT) < 0.01$. 
The second key point is the trend of the reversion temperature $T_{rev}$ with $E$ : using Eq. (17), i.e., $\beta=1/2$,  
leads 
$T_{rev}$ to increase by a factor three when $E=30 \times E_{rev}$, where $E_{rev}$ is the electrical field such that the 
nonlinearities onset at $T_{rev}$. This is in good agreement with Fig. 3 of Ref. \cite{Rogge}. On the contrary, using 
$\beta=1$ 
leads $T_{rev}$ to increase much faster with $E$ : $T_{rev}(E=30 \times E_{rev}) = 30 \times T_{rev}(E=0)$. Finally, the 
key 
role of $\beta=1/2$ is somehow reminiscent of Eq. (9) where $\delta \chi' \propto \sqrt{E}$, even if an analytical argument 
supporting this idea is still lacking.

With respect to experimental data, a failure, at this step of the discussion, is the ratio between the two slopes $\partial 
\epsilon'_r/\partial \ln{T}$ below and above the reversion temperature. In Fig. 3 this ratio is near -1.7:1 instead 
of -1:1 in most experiments.  Furthermore, the low-temperature
\textit{linear}-susceptibility data tend to a $T$-independent plateau while they do not in our simulations.  At very low 
temperature, interactions are likely 
to be so strong that the independent TLS model does not apply anymore, even with a renormalized relaxation time such as 
that 
of Eq. (17).  A transition toward a dipole-glass was invoked to explain the behavior of the samples whose $\chi'$ no longer 
depends on $T$ below a few mK. In this picture,
dipole orientation is progressively frozen, which would lead to a
plateau of the susceptibility \cite{Enss},\cite{Wurger} : by continuity, this would weaken the slope ratio near -1:1. 
Since the TLS model should not apply at very low $T$, it is not surprising that the plateau of the susceptibility measured 
in 
the nonlinear regime is not well accounted for by Fig. 3. Indeed, Fig. 3 does not show a completely $T$-independent plateau 
but only a substantial reduction of the $T$-dependence of $\chi'$ at low $T$: as stated in {\bf II)B)}, this is due to 
$\chi'_x$ 
which still exhibits a small $T$ dependence, even when $\chi'_z$ has turned into its $T$ independent regime. However, if, 
on Fig. 3, 
the susceptibility is 
frozen below a given $T$, one gets plateaus for $\chi'$ whose heights depend on $E$, as in experiments. Finally, pushing 
$\beta$ toward $1$ strengthens the tendency of $\chi'$ to become $T$ independent at low $T$ (unreported), even if $\beta 
\simeq 1$ leads to the above-mentioned discrepencies with respect to experimental data.
 Let us note that some materials (see Rogge \textit{et al.} \cite{Rogge}) do not yield any sign of such a glass transition 
even at $T=0.6$ mK.

\subsection{New predictions}

Let us move briefly to the physical predictions implied by Burin \textit{et al.}'s mechanism. Remembering that the 
inequality 
$\omega \tau_2 \ll 1$ allowed the key simplification for the derivation of $\chi'(E,T)$ -see Eq. (5)- we restrict ourselves 
to the 
kHz range where this condition is fulfilled. Two main predictions can be done : 

 \textit{i)} $\tau_{1B}(T)$ will be suppressed in samples whose thickness $h$ is smaller than the distance $\lambda_B$ 
separating the quasi-similar TLS's required by Burin \textit{et al.}'s mechanism. Indeed, at distances larger than $h$, 
dipolar interactions within the dielectric will be suppressed by the screening effect of the numerous electrons of the 
electrodes. Thus, if $h \lesssim \lambda_B$, one should observe a non linear behavior such as the one calculated in section 
{\bf I)} -see Fig. 1-, where the quantum saturation of the levels only remains. In other words, ranging $h$ from a fraction 
of 
$\lambda_B$ to a few $\lambda_B$ in a series of samples and studying  $\chi'(E,T)$ should lead to a gradual transition from 
Fig. 1 to Fig. 3 if Burin \textit{et al}'s mechanism is relevant, while it should not affect the non linear behavior in the 
standard 
TLS model. Note that such an experiment looks feasible due to the quite large value of $\lambda_B \simeq 10$ nm, -see {\bf 
II)A)}-. This is due to the fact that Burin \textit{et al.}'s mechanism requires the two interacting TLS's to have both 
very 
close values of $\Delta$ and very close values of $\Delta_0$ : these conditions are stringent enough to make $\lambda_B$ 
much 
larger than the distance between a given TLS and its nearest neighbor. 

\textit{ii)} The net relaxation frequency $\tau_{1}^{-1}+ \tau_{1B}^{-1}$ of a given TLS increases as $E$ increases. Thus, 
\textit{nonequilibrium data should be of smaller amplitude when $E$ is raised}. Indeed, they are currently interpreted as 
resulting from the very large $\tau_{1}$ existing in any glass due to the subclass of TLS's whose energy barrier is so high 
that $\Delta_0$ is very small. These very "slow" TLS's have an extremely delayed response to any change of the external 
constraints, such as the d.c. electrical, or strain, field imposed to the sample: these TLS's yield an excess of states at 
low 
energy with respect to the equilibrium density of states, the latter having a small depression at low energies due to 
TLS-TLS 
interactions. To our knowledge, the influence of $E$ on nonequilibrium phenomena has been reported only once, in Rogge 
\textit{et al.}'s work devoted to nonequilibrium phenomena on a mylar sample \cite{Rogge2}. Applying a  relative strain 
field 
$\cal F$ to the sample leads to a sudden jump of the dielectric capacity $C$, measured at $5$ $\mathrm{kHz}$, followed by a 
logarithmic relaxation. At $T=11$ $\mathrm{mK}$, i.e., well below $T_{rev}$, and with ${\cal F}=2.7\times 10^{-6}$, the 
initial relative jump is $dC/C = 13\times 10^{-7}$ if the measuring field is $E= 5\times 10^{4}$ $\mathrm{V/m}$ (see Fig. 1 
of 
Ref. \cite{Rogge2}), while it \textit{decreases} to $dC/C = 4.5\times 10^{-7}$ if the measuring field is $E= 8.5\times 
10^{4}$ 
$\mathrm{V/m}$ (see Fig. 2 of Ref. \cite{Rogge2}). Let us note that, with $p_0=1$ $\mathrm{D}$ and a relative dielectric 
constant 
of $5$, $E= 5\times 10^{4}$
 $\mathrm{V/m}$ amounts to an energy of $10$ $\mathrm{mK}$, of the order of $T$ : in terms of our Fig. 3 this means that 
one 
stands just above the pseudolinear regime, i.e., in a regime where our calculations, as well as Burin's mechanism, should 
apply. Even if this was not investigated systematically, this single experimental datum favors the idea that nonequilibrium 
effects should be of smaller amplitude when $E$ is increased, due to the interaction-induced reduction of the diagonal 
relaxation time. 

\section{Conclusions } 
In conclusion, we have simulated the nonlinear dielectric susceptibility of amorphous materials by using the TLS
model.  Phase coherence effects have been taken into account, which is the main difference with the adiabatic 
approximation.  
 In the kHz range, the standard TLS model yields a nonlinear behavior at odds with experiments due to the field induced 
depression of the quantum response. However, it was possible to fit in many details the 
experimental low-temperature
field-induced rising response by adding a new relaxation mechanism based upon the existence of 
interactions
below 100 mK.  In this approach, the low temperature response mainly loses its quantum origin at low frequency. Our work 
stresses the 
necessity 
to
inject interactions into the TLS model to get satisfactory predictions.

\acknowledgements{Many thanks to P. Pari, P. Forget, P. Ailloud (CNRS/LPS) and P. Trouslard (CEA/INSTN/LVdG)  
for help in experiments motivating this theoretical work. Scientific dicussions with J. Joffrin and J.-Y. Prieur (CNRS, Orsay) 
turned out to be crucial for this work, as well as the indication of Ref. \cite{AtomPhoton} by Pf. O. H. Rousseau (University 
of Perpignan).  
Useful discussions with D. Boutard, M. Ocio, M. Rotter, 
 are also acknowledged.}

\section{Appendix}

\subsection{Phase decoherence induced by small TLS interactions}

In this Appendix, we aim at giving some physical insight into the relaxation term introduced in the dynamics of an ensemble 
of 
TLS's due to their small mutual 
interactions. Expanding on the assumption that these interactions are much
 smaller than the other relevant energy scales (such as $T$ or the gap $\epsilon$),
 the basic idea \cite{Abragam} is to model these interactions by a small 
random electric field acting on each TLS. This idea is not new \cite{Halperin}, \cite{Abragam}, and numerical results are 
presented here only to help  understand theoretical results.

\subsubsection{Interactions effects when the measuring field $\mathbf{E}=0$}

Consider first the case where the measuring field $\mathbf{E}=0$. Modeling mutual interactions between TLS's by a random 
electric field leads, for a given TLS, to a total Hamiltonian given, by :

$$H = \frac{1}{2} \left( \begin{array}{cc}\epsilon&0\\0&-\epsilon\end{array} \right) + \left( 
\begin{array}{cc}\frac{\Delta}{\epsilon}&\frac{\Delta_0}{\epsilon}  \\ 
\frac{\Delta_0}{\epsilon}&-\frac{\Delta}{\epsilon}\end{array} \right) \mathbf{p_0}.\mathbf{E}_{rand} , \eqno(A1) $$

where the electric field $\mathbf{E}_{rand}$ is random in time for the considered TLS, and, at a given instant $t$, varies 
randomly for various TLS's. Note that Eq. (A1) is expressed in the eigen basis of the TLS.

Defining the density operator $\rho(t)$ by : 
$$\rho(t) = \left( \begin{array}{cc}\frac{1}{2}+z&x+iy\\x-iy&\frac{1}{2}-z\end{array} \right) , \eqno(A2) $$

it is clear that $x,y,z$ are, respectively, the quantum mean values of the three spin operators ($\bar{S_{x}},\bar{S_{y}}, 
\bar{S_{z}}$ are the corresponding symbols once the ensemble average over many similar TLS's is made).
 By using $i\hbar 
\dot{\rho}= H\rho - \rho H$, where the dot stands for time derivation, 
the dynamics of $x,y,z$ follows : 
$$\left\{
\begin{array}{cccccr}
\dot{z}&=&-\Omega_{1}y& \hbox{,}&\hbar\Omega_{1}=-2{\frac{\Delta_{0}}{\epsilon}}p_{0} E_{rand}&(A3a)  \\
\dot{x}&=&-\Omega_{0}y& \hbox{,}&\hbar\Omega_{0}=\epsilon + 2{\frac{\Delta} {\epsilon}}p_{0} E_{rand}&(A3b) \\
\dot{y}&=&\Omega_{0}x + \Omega_{1}z& & &(A3c) \\
\end{array}
\right. . $$

To characterize the random fluctuations in time of $E_{rand}$ we model its autocorrelation function by 
$<E_{rand}(t)E_{rand}(t+t')>_{t} = {\frac{u^2}{p_{0}^{2}\tau_{c}}}[\theta(t'+\tau_{c})-\theta(t'-\tau_{c})]$ where 
$\theta(t)$ 
stands for the Heaviside step function, $\tau_{c}$ is the characteristic time scale of the fluctuations and 
${u}/\sqrt{\tau_{c}}$ the typical scale of the fluctuating part of the Hamiltonian $H$. This means that $E_{rand}(t)$ is 
drawn 
at random once every $\tau_{c}$ and 
can be considered constant over time intervals 
$[n\tau_{c}, (n+1)\tau_{c}]$, where $n$ is an integer. Within each of these intervals, $E_{rand}(t)$ takes the constant 
value 
$E_{n}$. This allows to solve exactly the equation for $\ddot{y}$ obtained from Eqs. (A3) : $\ddot{y} + 
(\Omega_{0,n}^{2}+\Omega_{1,n}^{2})y=0$. This yields :

$$y(n\tau_{c} + t)= y(n\tau_{c}) \cos\Omega_{n}t + 
{\frac{\dot{y}(n\tau_{c})}{\Omega_{n}}} \sin\Omega_{n}t , \eqno(A4)$$

where $\Omega_{n}= \sqrt{\Omega_{0,n}^{2}+\Omega_{1,n}^{2}}$ with
 $\Omega_{0,n}$ and $\Omega_{1,n}$ defined as in Eqs. (A3) by setting $E_{rand}(n \tau_{c}+t)=E_{n}$. Inserting Eq. (A4) 
into 
Eq. (A3a) and Eq. (A3b), with the 
notation $X_{n} = X(n\tau_{c})$ for any quantity $X$, we get : 
$$\left\{
\begin{array}{cccr}
x_{n+1}&=&x_{n} - {\frac{\Omega_{0,n}y_{n}}{\Omega_{n}}}s_{n}-{\frac{\Omega_{0,n}\dot{y}_{n}}{\Omega_{n}^{2}}}(1-c_{n}) 
&(A5)  
\\
z_{n+1}&=&z_{n} - {\frac{\Omega_{1,n}y_{n}}{\Omega_{n}}}s_{n}-{\frac{\Omega_{1,n}\dot{y}_{n}}{\Omega_{n}^{2}}}(1-c_{n}) 
&(A6) 
\\
\dot{y}_{n+1}&=& \Omega_{0,n+1}x_{n+1}+ \Omega_{1,n+1}z_{n+1}&(A7) \\
\end{array}
\right. ,  $$

where $s_{n} = \sin\Omega_{n}\tau_c$, $c_{n} = \cos\Omega_{n}\tau_c$. 
The four equations (A4)-(A7) allow to deduce $x,y,z$ at step $(n+1)$ provided the corresponding quantities are known at 
step 
$n$. Choosing the
 initial conditions $x_{1},y_{1},z_{1}$, yields $\dot{y}_{1} = \Omega_{0}x_{1} + \Omega_{1}z_{1}$ which allows to initiate 
the 
recurrence. Finally, let us note that choosing the initial quantum state as $\vert \Phi_{1}> = a_{1} \vert +> + 
\sqrt{1-\vert 
a_{1} \vert^2} 
\exp(i \varphi_{1}) \vert - >$, where $\vert +>, \vert ->$ are the eigen states of the TLS, amounts to setting : $x_{1} = 
\vert a_{1} \vert \sqrt{1-\vert a_{1} \vert^2} \cos\varphi_{1}$, $y_{1} = \vert a_{1} \vert \sqrt{1-\vert a_{1} \vert^2} 
\sin\varphi_{1}$, $z_{1} = \vert a_{1} \vert^2 -1/2$.

\begin{figure}
 \includegraphics[height=6.5cm, width=8cm]{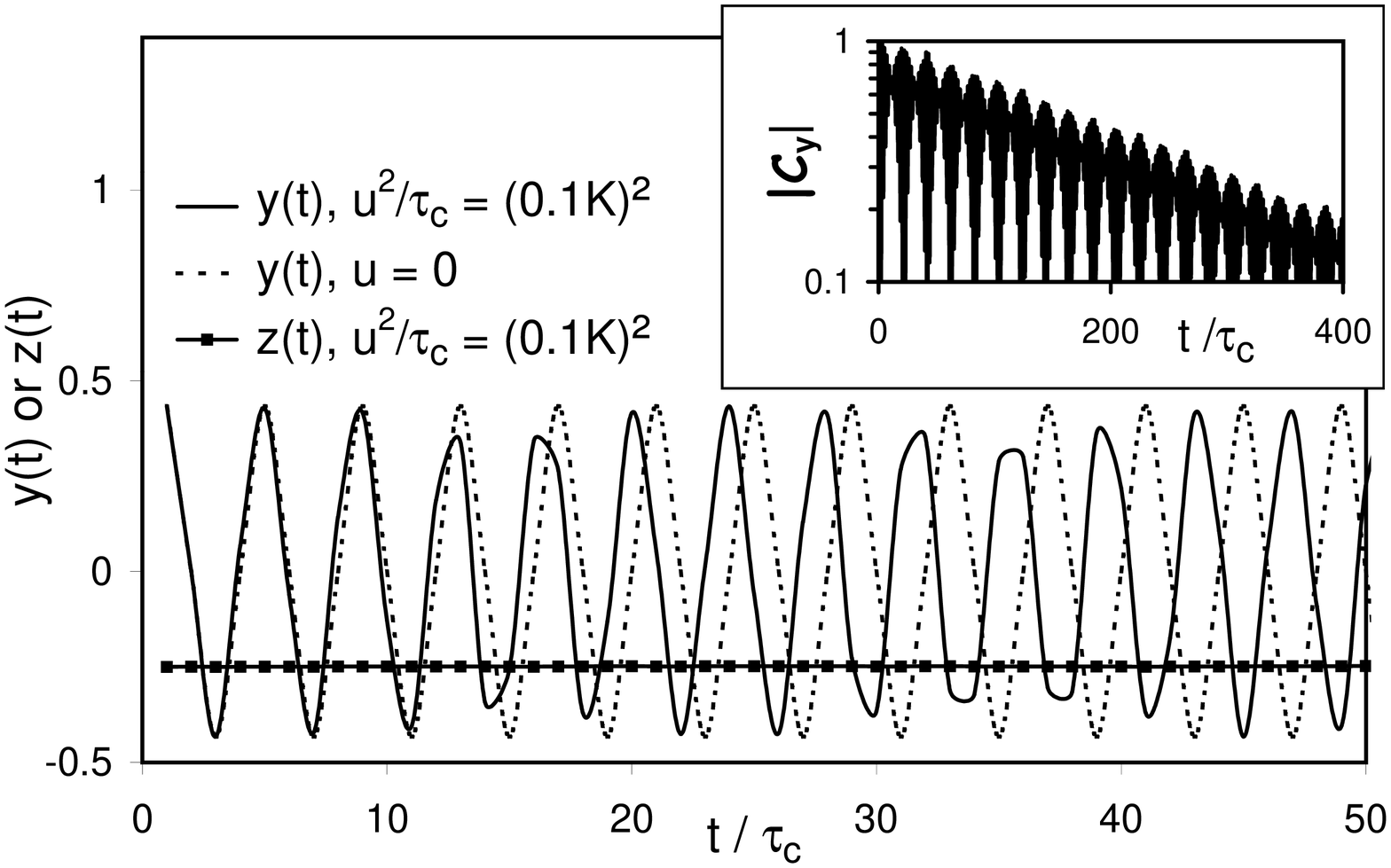}
 \caption{Dynamics of a TLS ($\Delta=1$ $\mathrm{K}$, $\Delta_{0}=0.01$ $\mathrm{K}$) submitted to a random electric field 
($u/\sqrt{\tau_{c}} = 0.1$ $\mathrm{K}$, $\tau_c$ is the quarter of the Bohr period $h/\sqrt{\Delta^{2} + 
\Delta_{0}^{2}}$). 
$z$, the quantum mean value of $S_{z}$, is basically constant (solid line with square symbols), i.e., mostly unchanged by 
the 
random electric field. On the contrary, $y$, the mean quantum value of $S_{y}$,
 is strongly affected by random electric field : the periodic Bohr oscillations (short dashed line) seen in the absence of 
random electric field, are progressively distorded when random electric field is present.\textit{ Inset:} As a result, 
${\cal 
C}_y$, the normalized autocorrelation function of $y(t)$, decreases exponentially with time.}
 \label{Fig. 4}
 \end{figure}

Figure 4 shows the dynamics of a TLS defined by $\Delta= 1$ $\mathrm{K}$, $\Delta_{0} = 0.01$ $\mathrm{K}$ evolving from 
the 
initial state $a_{1} = 1/2; \varphi_{1} = \pi/2$, i.e., from $x_{1} = 0; y_{1} = \sqrt{3}/4; z_{1} = -1/4$. The random 
field 
characteristics were set to $u/\sqrt{\tau_{c}} = 0.1$ K and $\tau_{c}=h/(4\epsilon)$, i.e., $\tau_{c}$ was chosen four 
times 
lower than the Bohr period. Without 'noise', $y(t)$ exhibits the well-known regular Bohr oscillations (short-dashed line on 
Fig. 4). 
The effect of 'noise'
 is to deform these ocillations (continuous line on Fig. 4) by an amount increasing with time : as a result the periodicity 
of 
$y(t)$ gradually disappears. This is illustrated in the inset of Fig. 4 showing the exponential decrease in time 
of the absolute value  $\vert {\cal C}_{y} \vert$ of the autocorrelation of $y$, 
defined by  
${\cal C}_{y}(t) = <\delta y(t') \delta y (t'+t)>_{t'}/\lambda^2$ with 
 $\delta y (t) = y(t)-<y>$ and $\lambda^{2}= <(\delta y)^{2}>$.

 Since 
the value $y_{n}$ depends on the set of values $E_{n}$ drawn for the considered
 TLS from $n=1$, ensemble averaging (over many TLS's with the same $\Delta, \Delta_{0}$) will lead to a cancelation of $y$ 
due 
to the absence of correlations between the noise series seen by different 
TLS's. This cancelation happens on a time scale $\tau_{2}$ which should be of
 the order of the one of ${\cal C}_{y}$ shown in the inset of Fig. 4. 
This cancelation of $y$ after ensemble averaging amounts to a 
supplementary relaxation 
term $\bar{S}_{y}/\tau_{2}$ in the Bloch equation describing $\bar{S}_{y}$ dynamics. 

The dynamics of $x(t)$ (unreported on Fig. 4) is similar to the one of $y$, yielding a corresponding relaxation term 
$\bar{S}_{x}/\tau_{2}$. 
This contrasts totally with the dynamics of $z(t)$, depicted on Fig. 4: provided the amount of noise $\delta H(t)$ is much 
smaller than the gap $\epsilon$, $z(t)$ stands very close to its initial value $z_{1}$, even at large times. In fact small 
fluctuations exist, with an autocorrelation decrease similar to the one of ${\cal C}_{y}$, but the key point is that $\vert 
z(t)/<z> -1\vert \ll 1$. Hence the 'noise' does not yield any supplementary relaxation term in the Bloch equation governing 
the population dynamics $\bar{S}_{z}$.

\subsubsection{Interaction effects with a finite measuring field $\mathbf{E}$.}

When the measuring $\mathbf{E}$ is no longer zero, the whole dynamics 
should be recalculated, with the supplementary dipolar Hamiltonian
 corresponding to $\mathbf{E}$. However, the fact that the measuring 
frequency $\omega$ is much lower than $1/\tau_{2}$ greatly simplifies the problem. Indeed, if $\omega$ were zero, taking 
into 
account of $\mathbf{E}$ would strictly amount to replace $\Delta$ by $\Delta+ \mathbf{p}_{0}.\mathbf{E}$ : with this new 
definition of $\Delta$,  all the previous calculations apply, yielding the same relaxation terms in the Bloch equations. We 
will assume that this holds true for finite $\omega$, due to the fact that for the kHz frequencies considered here, the 
experimental values of $\tau_{2}$ ensure $\omega \tau_{2} \ll 1$, even at the lowest $T$ studied in the body of the paper.

\subsection{Validity of Bloch equations.}
The three Bloch equations Eqs. (3a)-(3c) are valid in the quasilinear response \cite{DeVoe}.  When the electric field 
becomes strong 
enough, 
the relaxation terms form a nondiagonal matrix, e.g.  a  $\bar{S_{z}}/\tau_{x,z}$ term might come into play in the first 
Bloch equation, and the corresponding Bloch equations are usually named in the litterature Generalised Bloch Equations 
(G.B.E.). However, up to our knowledge, these generalized relaxation terms have been calculated only in the case of 
transverse 
fields in the rotating wave approximation \cite{Geva}. This is at odds with our physical situation : i) The transverse 
field 
case amounts to $\Delta=0$, which, by far, is not the case considered here (remember that, due to the $1/\Delta_{0}$ 
density 
of states, for most TLS's one has $\Delta \ge \Delta_{0}$) ; 
ii) The measuring field $E \sim cos\omega t$ is an oscillating one, not a rotating one $\sim \exp i\omega t$ and the 
rotating 
wave approximation would be
 valid only close to the resonance $\omega \simeq \epsilon/\hbar$, a condition totally irrealistic here due to the extreme 
smallness of $\hbar \omega = 2 \times 10^{-7}$ K.

However, even if they do not apply in our case, one can use the G.B.E. derived in the rotating wave 
approximation for transverse fields to guess qualitatively what could be the influence of the off-diagonal relaxation 
terms. 
Two points are worth mentioning : 

\textit{i)} One can easily check that the GBE still yield qualitatively the quantum saturation phenomenon, even if the 
off-diagonal relaxation terms are responsible for quantitative modifications. In particular, it was shown, in the limit of 
infinite $E$, that the GBE reduce to the standard Bloch equations with $\tau_{2} = 2 \tau_{1}$ and that one gets a 
vanishing 
susceptibility. 

\textit{ii)} In the GBE, the off-diagonal relaxation times become infinite 
(i.e. negligible) when $\tau_{c} \to 0$, where $\tau_{c}$ is the correlation time of the random field created, on a given 
TLS, 
by its neighbors. In the same spirit \cite{Eberly}, in the GBE, $\tau_{2}$ is affected by  a multiplicative factor 
$\left(1+\Xi^2\tau_c^2\right)$ where
$\Xi=\left|\mathbf{p_0}.\mathbf{E}\right|/\hbar$ is the Rabi frequency. The order of magnitude of $\tau_c$ in glasses was 
measured only once by Devaud and Prieur \cite{Prieur} who found $\tau_{c} \simeq 10^{-8}$ s  at $T=70$ mK with an expected  
$\tau_c\sim 1/T$ temperature dependence.  The $E$ dependence in the
relaxation times can be neglected if $\Xi\tau_c \le 1$. Aware of these limits, we guess the standard Bloch equations can 
give 
a fair
approximation as long as $\left|\mathbf{p_0}.\mathbf{E}\right|$ does not exceed $0.1-1$ mK at low temperature. As an 
additional remark, the validity domain of our calculations extends as $\tau_{c}$ decreases.

To summarize, the GBE do not suppress the quantum saturation phenomenon, on the contrary, they are intended to 
quantitatively 
account for the various measurable quantities in the saturation regime (such as linewidths, etc...). The problem of the 
strong 
depression of $\chi'$ when $p_0E$ is increased from extremely small values up to $10^{-4}-10^{-3}$ K is thus unavoidable 
and 
is at odds with Rogge \textit{et al.}'s experiments \cite{Rogge} which were carried out on \textit{various} glasses and 
showed  \textit{absolutely no sign} of field induced depression of $\chi'(T<T_{rev})$, despite the fact that $p_0E$ was 
varied 
from $0.05$ mK to $50$ mK : the fact that the domain $p_0E \le 1$ mK was 
experimentally investigated is of special importance since, as stated above, in this domain, at least, the Bloch equations 
used here should be valid.

%\bibliography{prb2}

\begin{thebibliography}{10}
\bibitem{noteadia}
\bibinfo{journal}{This contrasts with the dissipative part $\chi''$ of the susceptibility where the adiabatic approximation seems 
more efficient, which is not surprising since it is well established that both parts of $\chi$ are \textit{not} dominated by the same 
kind of TLS's }.
\bibitem{t1t}
\bibinfo{journal}{The other underlying key assumption is that the phonon density remains at
equilibrium when a phonon is emitted or absorbed.  This assumption may fail if the time for the emitted phonon to be 
thermalized is not
negligible, inducing a local nonequilibrium phonon density called "bottleneck effect" \cite{Scott}.  However, this effect 
has 
not been observed
in glasses \cite{Leggett}, suggesting $\tau_1$ is indeed time-independent}.
\bibitem{2t1}
\bibinfo{journal}{In the case of strictly non interacting TLS's, $\tau_2$ becomes infinite and the relaxation of both $S_x$ 
and $S_y$ 
is only ensured by the $S_{x,y}/2\tau_1$ terms. }.
\bibitem{notefin}
   \bibinfo{journal}{The $E^{3/4}$ exponent is not exact since, in the Fourier Transform of Eq. (18), we neglected the fact 
that  
$\tau_z/\tau_{z,1}$ is time dependent and that all its harmonics couple to $<\bar{S}_z^n>$ to contribute to $\bar{S}_z^{1}$ 
.}
\end{thebibliography}

\end{document}